\documentclass[journal]{IEEEtran}
\usepackage{amsmath,amsfonts}
\usepackage{algorithmic}
\usepackage{algorithm}
\usepackage{textcomp}
\usepackage{stfloats}
\usepackage{url}
\usepackage{verbatim}
\usepackage{graphicx,wrapfig}
\usepackage{svg}
\usepackage{cite}
\usepackage{picinpar}
\usepackage{amssymb}
\usepackage{esvect}
\usepackage{flushend}
\usepackage[latin1]{inputenc}
\usepackage{colortbl}
\usepackage{soul}
\usepackage{multirow}
\usepackage{pifont}
\usepackage{color}
\usepackage{xcolor}
\usepackage{alltt}
\usepackage[hidelinks]{hyperref}
\usepackage{enumerate}
\usepackage{siunitx}
\usepackage{epstopdf}
\usepackage{pbox}
\usepackage{booktabs}
\usepackage{tabularx}
\usepackage{array}
\usepackage{cite}
\ifCLASSOPTIONcompsoc
    \usepackage[caption=false, font=normalsize, labelfont=sf, textfont=sf]{subfig}
\else
\usepackage[caption=false, font=footnotesize]{subfig}
\fi

\newtheorem{proposition}{\bf{Proposition}}[section]
\newtheorem{proof}{\bf{Proof}}[section]
\newtheorem{remark}{\bf{Remark}}

\begin{document}

\title{Enhanced Grid Following Inverter (E-GFL):\\ A Unified Control Framework for Stiff and Weak Grids}

\author{
	\vskip 1em
	
	Alireza Askarian\textsuperscript{1,a}, \emph{Student Member}, 
	Jaesang Park\textsuperscript{1,b}, \emph{Student Member},
	\\ and Srinivasa Salapaka\textsuperscript{1,c}, \emph{Senior Member}

	\thanks{
	
		\textsuperscript{1} Department of Mechanical Science and Engineering, University of Illinois at Urbana-Champaign, 61801 IL, USA
		
		\textsuperscript{a}askaria2@illinois.edu,\textsuperscript{b}jaesang4@illinois.edu,\textsuperscript{c}salapaka@illinois.edu
		
		The authors would like to acknowledge the support of Advanced Research Projects Agency-Energy (ARPA-E) for supporting this research through the ARPA-E OPEN project titled "Rapidly Viable Sustained Grid" via grant no. DE-AR0001016.
	}
}

\markboth{Journal of \LaTeX\ Class Files,~Vol.~14, No.~8, August~2021}%
{Shell \MakeLowercase{\textit{et al.}}: A Sample Article Using IEEEtran.cls for IEEE Journals}


\maketitle

\begin{abstract}
This paper presents an extensive framework focused on the control design, along with stability and performance analysis, of Grid-Following Inverters (GFL). It aims to ensure their effective operation under both stiff and weak grid conditions. The proposed framework leverages the coupled algebraic structure of the transmission line dynamics in the $dq$ frame to express and then mitigate the effect of coupled dynamics on the GFL inverter's stability and performance. Additionally, we simplify the coupled multi-input, multi-output (MIMO) closed-loop system of the GFL into two separate single-input, single-output (2-SISO) closed-loops for easier analysis and control design. We present the stability, robust stability, and performance of the original GFL MIMO closed-loop system through our proposed 2-SISO closed-loop framework. This approach simplifies both the control design and its analysis. Our framework effectively achieves grid synchronization and active damping of filter resonance via feedback control. This eliminates the need for separate phase-locked loop (PLL) and virtual impedance subsystems. We also utilize the Bode sensitivity integral to define the limits of GFL closed-loop stability margin and performance. These fundamental limits reveal the necessary trade-offs between various performance goals, including reference tracking, closed-loop bandwidth, robust synchronization, and the ability to withstand grid disturbances. Finally, we demonstrate the merits of our proposed framework through detailed simulations and experiments. These showcase its effectiveness in handling challenging scenarios, such as asymmetric grid faults, low voltage operation, and the balance between harmonic rejection and resonance suppression.
\end{abstract}


\vspace{-5mm}
\section{Introduction}
\begin{figure}[t]
    \centering
    \subfloat[]{\includegraphics[width = 0.75\linewidth]{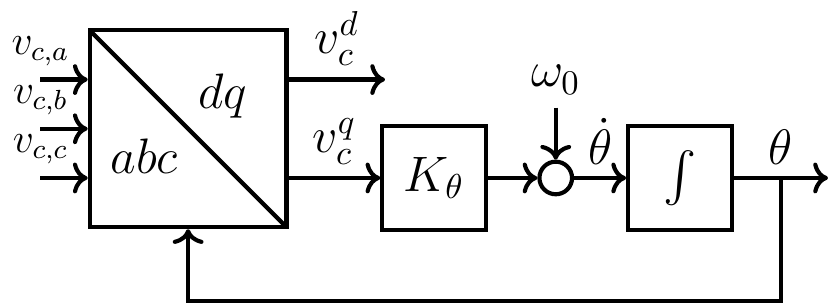}
    \label{fig:PLL_Conventional}}
    \\
    \vspace{-4mm}
    \subfloat[]{\includegraphics[width = 0.75\linewidth]{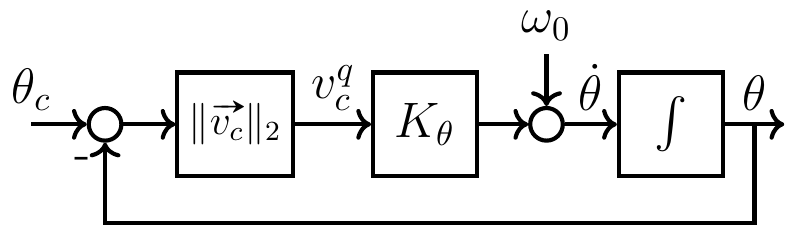}
    \label{fig:PLL_Linearized}}
    \caption{(a) Conventional 3-phase PLL system that include $abc$ to $dq$ transform, PLL loop filter $K_\theta$ and voltage controlled oscillator (VCO) (integrator). (b) Linearized model of the PLL in (a) that is frequently used for stability and performance analysis of the PLL.}
\end{figure}

\begin{figure}
    \centering
    \begin{tabular}{c@{}c}
    \subfloat[]
    {\includegraphics[width = 0.6\linewidth]{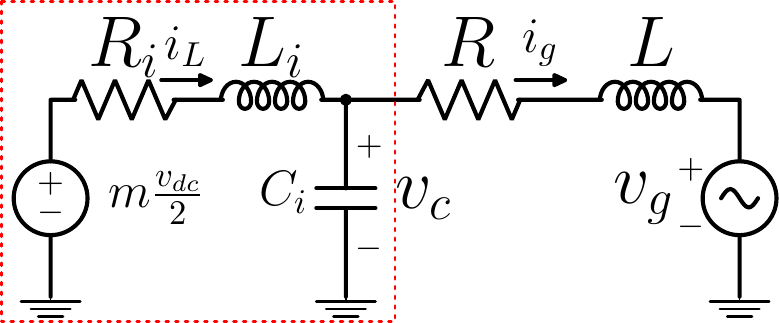}
    \label{fig:Averaged_Inverter_and_Line}}
    &
    \subfloat[]
    {\includegraphics[width = 0.36\linewidth]{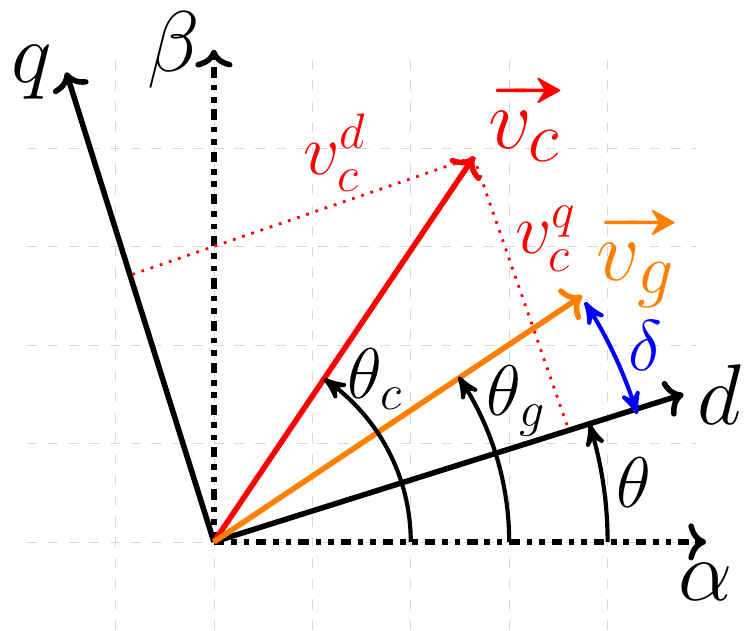}
    \label{fig:DQ_Frame}}
       
    \end{tabular}
    \caption{(a) Inverter averaged model with output $RLC$ filter, interfaced to the grid via $RL$ impedance. (b) The $dq$ rotating frame.}
    \label{fig:my_label}
    \vspace{-5mm}
\end{figure}

\IEEEPARstart{I}{n} the last two decades, there has been a significant effort to develop a decentralized power network that enables the large-scale integration of distributed energy resources (DER) \cite{perez2021roadmap}. This has led to the development of {\em microgrids}, where power subnetworks are formed to establish an autonomous grid or connect to the main power grid. Switching power systems, with their powerful digital processing capabilities and fast dynamics, are the prime enablers of the distributed power framework. Among switched power systems, DC/AC inverters are the most widely adopted solution to interface DERs with the AC power network. DC/AC inverters provide a flexible platform to inject active and reactive power into the grid in the grid-following (GFL) mode or provide voltage and frequency support at the point of common coupling (PCC) in the grid-forming (GFM) mode. The primary focus of this work is on the GFL mode of operation. However, many of the proposed concepts can be readily adapted to GFM operation with little or no change.\\  
The main function of the GFL inverter is to track a predetermined power set point. This power set point generally reflects specific economic objectives related to power efficiency, such as tracking the maximum power point in photovoltaics (PV) or providing ancillary services to the power grid \cite{zarei2019control,blaabjerg2017distributed,yuan2021real}. However, maintaining an adequate stability margin and satisfactory performance is often challenging due to the undesired dynamical artifacts caused by nonlinearities, switching distortion, and uncertainties in the system. Furthermore, grid disruptions, which are common in weak grids and microgrids, only exacerbate these difficulties. Therefore, operating a GFL inverter connected to a weak grid involves several key research focuses:
\\
$\bullet$ \textbf{Damping the Filter Resonance:} A major problem with the GFL inverter is the considerable resonance that arises in inverter and line dynamics. This is mainly due to the under-damped modes of the inverter filter, particularly the $LCL$ filter.
\\
$\bullet$ \textbf{Mitigating Harmonics:} Nonlinear loads, interactions with the weak grid, and switched devices can all contribute to harmonics of fundamental frequency. It is essential to suppress these harmonics in the current waveform to ensure that the quality of the output current waveform is maintained.
\\
$\bullet$ \textbf{Reliable Synchronization Scheme:} 
Effective operation of the GFL inverter requires a synchronization scheme that can lock on to the grid voltage and maintain stability and phase lock, when faced with grid disturbances such as sudden changes in voltage/frequency, abrupt phase shifts, and voltage imbalances.
\\
$\bullet$ \textbf{Power Tracking and Stability:} Under weak grid conditions, it is challenging to ensure effective power tracking and stability. Variability in system parameters, such as line impedance, which often occurs in weak grid scenarios, requires a robust approach to stability.
\\
$\bullet$ \textbf{Grid Fault Ride-Through:} Demonstrating the ability of the inverter to withstand and operate through grid faults is critical to overall reliability and resilience.
\\
These areas are central to improving the reliability and efficiency of GFLs in weak grid environments. In what follows, we will offer a concise overview of the aforementioned topics.
\\
Resonance, whether caused by the underdamped $LCL$ filter or the interaction between the inverter's $LC$ filter and the inductive transmission line, can lead to instability. Active damping techniques provide a promising solution to mitigate filter resonance and improve stability, without relying on loss-inducing components \cite{wang2015grid,bao2013step,he2020hybrid,chen2023unified,wang2021passivity, awal2021observer,zhao2021passivity}. Using virtual impedance to dampen the resonance of the $LCL$ filter is a method that is intuitive, efficient, and easy to implement. This technique utilizes feedback measurements from the filter states, such as capacitor voltage or inductor current, to imitate the effects of a loss-inducing circuit element, such as a resistor, on the inverter's output voltage and current. The virtual impedance technique has been proved to be effective in enhancing the damping characteristics of the filter and the stability margins of the GFL inverter \cite{wang2015grid, bao2013step,he2020hybrid,chen2023unified}. More recently, passivity-based approaches to active damping have gained popularity \cite{wang2021passivity, awal2021observer, zhao2021passivity, ma2023passivity}. Passivity-based active damping is rooted in the concept of a dissipative system within control theory. This approach typically guarantees stability and robustness under a variety of operating conditions and parameter changes, which are common in weak grid scenarios. Passivity-based damping methods, though more complex than virtual impedance, offer a more comprehensive analytical framework and advanced damping schemes\cite{zhao2021passivity, ma2023passivity}. Although active damping methods are effective in reducing filter resonance, they can interfere with the functioning of the current closed-loop and Phase-Locked Loop (PLL). If not adjusted correctly, these methods can have a detrimental effect on closed-loop bandwidth, synchronization quality, low-frequency disturbance rejection, and the ability to track reference power in steady state. \cite{bao2013step}.\\
Grid codes \cite{photovoltaics2018ieee,bajaj2020grid} require GFL inverters to suppress harmonics of the fundamental frequency and inject a distortion-free current into the grid. This is generally achieved using multifrequency proportional resonance (PR) controllers \cite{hans2020design, ye2016analysis, xin2016grid}. Theoretically, high-gain PR compensators provide better harmonic attenuation and a faster transient response, while increasing the PR damping reduces the attenuation factor but improves the PR sensitivity to changes in harmonic frequency \cite{hans2020design, ye2016analysis, xin2016grid}. The systematic PR control design is usually based on the harmonic attenuation factor and the phase margin specification at the corresponding harmonic frequency \cite{ye2016analysis, xin2016grid}. Although this approach provides the basis for tuning the PR gain and damping, it does not capture the interaction between the PR compensator, weak grid, and the inverter filter. This can cause resonance and instability for high values of PR gain and damping factor. Therefore, PR designs include a second step in which the closed-loop transient is tuned through trial and error on both the experimental and simulation platforms \cite{vidal2012assessment,husev2019optimization}.\\
Most GFL control designs employ a variation of the synchronous reference frame phase-locked loop (SRF-PLL), as shown in Fig. \ref{fig:PLL_Conventional}, to synchronize with the grid voltage \cite{xu2021overview}. The traditional SRF-PLL, depicted in Fig. \ref{fig:PLL_Conventional}, originates from the field of communication systems \cite{golestan2021frequency}. In this design, it is implicitly assumed that the phase angle of the input signal, denoted $\theta_c$ in Fig. \ref{fig:PLL_Linearized}, is independent of the phase angle $\theta$ generated by the SRF-PLL. In typical GFL applications, this assumption is only valid when the inverter is connected to a stiff grid with negligible line impedance \cite{shakerighadi2020modeling,islam2021accurate,xu2021analyses,9690060,ahmed2020linear}. Under this condition, the voltage phase angle remains dynamically separate from the SRF-PLL angle. However, in the presence of a weak grid and substantial line impedance, the input phase angle $\theta_c$ in Fig. \ref{fig:PLL_Linearized} becomes strongly coupled to the PLL dynamics and phase angle $\theta$ \cite{wang2017unified}. Consequently, synchronization algorithms based on traditional SRF-PLL are sensitive to voltage fluctuations in the grid and unreliable when there are low-order filter resonance, harmonics, unbalanced, and weak grid conditions \cite{golestan2012dynamics}. Kulkarni et al. \cite{kulkarni2012analysis} suggest a technique for tuning the PLL parameters to balance the PLL synchronization bandwidth and the robustness to voltage unbalance and grid harmonics. Furthermore, Gonzalez and Lee \cite{gonzalez2012adaptive,lee2013novel} have used an adaptive notch filter to reduce the oscillations of the PLL phase angle in a polluted AC system. Shakerighadi \cite{shakerighadi2020modeling}, employed a nonlinear time-varying framework for assessing the stability of the SRF-PLL and proposed an adaptive solution to improve stability bounds under frequency and phase disturbances. Zahidul Islam \cite{islam2021accurate} proposed an adaptive Clarke transform (ACT) algorithm to accurately track the phase angle of the grid under unbalanced voltage and phase condition. In addition, there is a large body of literature investigating different variations of the Kalman filter to improve the quality of synchronization \cite{xu2021analyses,9690060,ahmed2020linear}.\\
Recently, the development of an inertial PLL (IPLL) has emerged as a promising method to enhance the transient response and stability margin of SRF-PLL \cite{tarraso2023grid}\cite{li2021inertia}. The IPLL incorporates a virtual swing equation into the standard SRF-PLL structure. This integration enhances the SRF-PLL's ability to handle frequency deviations more effectively during transient conditions. Finally, there is increasing research on power synchronization techniques for GFL operation \cite{mansour2022linear}. These techniques, commonly used in grid-forming inverters, are being explored as alternatives to traditional SRF-PLL in GFL operations.\\
Research on GFL control has been extensive, yet the existing literature typically deals with the aforementioned challenges separately, even though they are closely coupled or related and stem from the same underlying dynamical system. Furthermore, some popular design techniques, such as virtual impedance, SRF-PLL, and virtual inertia, involve reusing familiar concepts and components from different fields. Although this approach is intuitive, it can impose artificial restrictions on the inverter's dynamics and limit the peak performance.\\
The primary contribution of this paper is an innovative and comprehensive framework for GFL closed-loop modeling, stability and performance analysis, control design, and control implementation. This framework stands out from the existing literature because of its unified approach to tackling the GFL control issues. Essentially, we evaluate robust stability, zero steady-state tracking, harmonic attenuation, synchronization, frequency transients, inertial response, active damping of filter resonance, and the influence of coupled inverter dynamics on stability and performance using the core components of the closed-loop structure. The unified approach in the proposed framework allows us to determine the fundamental trade-offs and bottlenecks between the different control objectives and subsequently design the controllers to achieve the proper trade-off. The following outlines the structure of this paper and provides an overview of each section's primary focus and contribution.\\
Section \ref{sec:GFL_Problem_Setting}, outlines the key dynamic equations and the context for evaluating the GFL control objectives. In Section \ref{sec:GFL_Closed-Loop_Dynamics_-_A_New Perspective}, we propose the closed-loop structure in Fig. \ref{fig:Averaged_Inverter_and_Line} where the transmission line dynamics forms the {\em plant}, and the inverter dynamics is treated as an actuator. The primary challenge in the proposed model arises from the coupled multi-input, multi-output (MIMO) closed-loop dynamics. We utilize the algebraic structure of the closed-loop to characterize the underlying coupled MIMO system in terms of a nominal 2-SISO closed-loop with multiplicative MIMO perturbation. As we show, this formulation allows us to clearly isolate the effect of coupled dynamics on the GFL closed-loop. Section \ref{sec:Characterization_of_Stability_Performance} expresses the performance objectives, stability, and robust stability conditions for the proposed MIMO closed-loop using direct specifications on the nominal 2-SISO closed-loops derived in Section \ref{sec:GFL_Closed-Loop_Dynamics_-_A_New Perspective}. This section examines the influence of dynamical coupling on the stability and performance of the GFL inverter and provides a solution to reduce the effect of coupling on the closed-loop. Furthermore, this section introduces an innovative synchronization method that relies only on the disturbance rejection characteristics of the feedback controller and effectively omits the need for an additional PLL system. The proposed synchronization scheme offers a systematic approach to achieve inertia, robustness to grid disturbances, and high-frequency harmonic attenuation. Furthermore, we show that our synchronization technique remains effective even under asymmetrical grid fault. Section \ref{sec:water_bed} explores and quantifies the fundamental trade-off between stability margins and control objectives, such as reference tracking, grid disturbance rejection, and damping high-frequency resonance, all based on closed-loop sensitivity analysis. This section demonstrates that linear control design inevitably faces certain performance limitations. As a result, a well-founded control design relies on balancing these desired objectives. In Section \ref{sec:Framework_for_Inverter_Control}, we present a control design example that meets the stability and performance criteria established in this paper. In addition, we show how the closed-loop dynamics of the inverter can be integrated into the feedback controller to implement the proposed feedback controllers.
\\
In the simulation and experimental sections, we focus on key aspects of our proposed framework. The first experiment explores control designs that achieve different trade-offs between harmonic attenuation and PR-induced filter resonance. This test confirms the fundamental limitations on GFL performance, as discussed and quantified in Section \ref{sec:water_bed}. The second experiment evaluates the resilience and transient response of the proposed synchronization scheme in the presence of frequency anomalies common in weak grids. In the third part, we showcase the system's robustness against asymmetrical faults, demonstrating how the control design effectively counters the resulting harmonics and matches the asymmetrical fault ride-through capabilities of the Decoupled Double Synchronous Reference Frame PLL (DDSRF). The fourth part examines the robust stability of our controller, based on the novel criteria from Section \ref{sec: SISO_Robust_Stability}. Here, we compare its stability margins with popular active damping techniques that use capacitor voltage and grid current feedback and show improvement in the robust stability margin. Finally, we evaluated the low voltage ride-through capability (LVRT), showing that the system operates reliably even when the grid voltage drops to 30\% of its nominal value.\\
The proof of the main results, presented in the form of propositions, can also be found in the appendix.
\vspace{-3mm}
\section{GFL Problem Setting}
\label{sec:GFL_Problem_Setting}
We study the GFL control in the context of two connected voltage sources shown in Fig. \ref{fig:Averaged_Inverter_and_Line}. Here, $v_c$ is the inverter's output capacitor voltage, and $v_g$ denotes the grid voltage. The $RL$ impedance between $v_c$ and $v_g$ represents the combined effect of the transmission line impedance, the grid impedance (for the weak grid), and the grid side inductor of the $LCL$ filter (if the $LCL$ filter is used). The controlled voltage source ${m}\frac{v_{dc}}{2}$ is the cycle-averaged model of the inverter switching node, where the modulation signal ${m} \in [-1,1]$ controls the output voltage of the switching node \cite{6739422,Erickson2020}. The inverter's output $LC$ filter is represented by $C_i$, $L_i$, and the equivalent series resistor (ESR) $R_i$. Furthermore, $i_L$ is the current that flows through $L_i$ and $R_i$. The line and inverter dynamics in Fig. \ref{fig:Averaged_Inverter_and_Line} is given by
{\small
\begin{align}
    L\frac{d{i_g}}{dt} &= {v_c} - {v_g} - R {i_g}, 
    \label{eq:line_dynamics_ab}
    \\
    \begin{split}
        L_i\frac{d{i_L}}{dt} &= \frac{v_{dc}}{2}{m} - {v_c} - R_i{i_L}, \quad
        C_i\frac{d{v_c}}{dt} = {i_L} - {i_g}.
        \label{eq:inverter_dynamics_ab}
    \end{split}
\end{align}}%
In this work, we adopt the direct-quadrature ($dq$) rotating frame to represent the above dynamics. The $dq$ frame allows us to (a) use the same notation, analysis, and design for one-phase and three-phase inverter systems, (b) transform the sinusoidal tracking problem into a dc tracking problem, (d) manipulate the phase directly, and (d) avoid nonlinear operations such as phase-shifting/scaling\cite{6739417} that are common in stationary frame control methods. In the $dq$ frame, the line and inverter dynamics in (\ref{eq:line_dynamics_ab}) and (\ref{eq:inverter_dynamics_ab}) are transformed into
{\small
\begin{align}
        L\frac{d\vv{i_g^{dq}}}{dt} &=
        \vv{v_c^{dq}} - \vv{v_g^{dq}} -
        R\vv{i_g^{dq}} -
        L\Dot{\theta}
        \begin{bmatrix}
        -i_g^q, i_g^d
        \end{bmatrix}^{\top},
    \label{eq:line_dynamics_ss_dq} 
    \\
    \begin{split}
    L_i\frac{d\vv{i_L^{dq}}}{dt} &=
    \frac{v_{dc}}{2}\vv{m^{dq}} - \vv{v_c^{dq}} -
    R_i\vv{i_L^{dq}} -
    L_i\Dot{\theta}
    \begin{bmatrix}
    -i_L^q, i_L^d
    \end{bmatrix}^{\top}, 
    \\
    C_i\frac{d\vv{v_c^{dq}}}{dt} &=
    \vv{i_L^{dq}} - \vv{i_g^{dq}} -
    C\Dot{\theta}
    \begin{bmatrix}
    -v_c^q, v_c^d
    \end{bmatrix}^{\top}.
    \label{eq:inverter_dynamics_ss_dq}
    \end{split}
\end{align}}%
In the above, $\Dot{\theta}$ denotes the angular frequency of the $dq$ rotating frame and each vector $\vv{f^{dq}}=[f^d,f^q]^{\top}$ comprises the corresponding $d$, and $q$ signal components as shown in Fig. \ref{fig:DQ_Frame} \cite{6739417}. In this paper, we primarily use the Laplace transform of the $dq$ dynamics in (\ref{eq:line_dynamics_ss_dq}) and (\ref{eq:inverter_dynamics_ss_dq}) given as
{\small
\begin{align}
    \begin{split}
        \vv{\hat{i}_{g}}
        &=
        G_L(s)
        \left(\vv{\hat{v}}_c - \vv{\hat{v}}_g\right),
    \label{eq:line_model}
    \end{split}\\
    \begin{split}
        \vv{\hat{v}_c} &= 
        G_v(s)
        \left(\vv{\hat{i}_L} - \vv{\hat{i}_g}
        - C_i \Dot{\theta} \left[-v_c^q,v_c^d\right]^{\top}
        \right),\\
        \vv{\hat{i}_L} &= 
        G_{i}(s)
        \left(\frac{v_{dc}}{2}\vv{\hat{m}} - \vv{\hat{v}_c}
        - L_i \Dot{\theta} \left[-i_L^q,i_L^d\right]^{\top} \right),
    \label{eq:inverter_ss_dynamics}
    \end{split}
\end{align}}%
where
{\small\begin{align}
    G_L(s) &= 
    \frac{
        \begin{bmatrix}
        s+\lambda & \omega_{0} \\
        -\omega_{0} & s+\lambda
        \end{bmatrix}}
        {L\left( s^2 + 2\lambda s + 
        \lambda^2 + \omega_{0}^{2}\right)},
        \;\;
        \lambda = \frac{R}{L},
        \label{eq:GL_expression}
        \\
        G_v(s) &= \frac{1}{C_i s},
        \;\; \text{{\normalsize and}} \;\;
        G_i(s)  = \frac{1}{L_i s + R_i}.
        \label{eq:Gv_Gi_expression}
\end{align}}%
We use the hat symbol to denote Laplace domain signals and omit the $dq$ superscript for vector signals. We assume that all vector signals are expressed in the $dq$ frame, unless otherwise specified.\\ 
Finally, the steady-state operating set-points for the GFL are typically specified in terms of power. However, to exploit the fast, rich, and linear dynamics of voltage and currents in (\ref{eq:line_model}) and (\ref{eq:inverter_ss_dynamics}), we map the power reference ($P_0$, $Q_0$) into the corresponding current set-points $\{i_0^d$, $i_0^q\}$ as
{\small\begin{equation}
    \begin{split}
        \begin{bmatrix}
        i_0^d \\ i_0^q
        \end{bmatrix} = 
        \frac{\Phi_{\{1, 3\}}}
        {\|\vv{v_c}(t)\|^2}
        \begin{bmatrix}
        v_c^d & \hphantom{-}v_c^q \\
        v_c^q & -v_c^d
        \end{bmatrix}
        \begin{bmatrix}
        P_0 \\ Q_0
        \end{bmatrix},
    \end{split}
    \label{eq:power_current_mapping}
\end{equation}}%
where $\Phi_1 = 1$ for single phase and $\Phi_3 = 2/3$ for three-phase system. The conventional PLL dynamics, such as that shown in Fig. \ref{fig:PLL_Linearized}, is responsible for forcing $v_c^q$ in (\ref{eq:power_current_mapping}) to zero by generating the proper rotating angle $\theta$ that aligns the rotating frame's $d$ axis with the inverter voltage phasor $\vv{v_c}$. This effectively decouples the mapping between the power reference $\{P_0,Q_0\}$ and corresponding current set-point $\{i_0^d,i_0^q\}$.
\begin{figure}[t]
    \centering
    \subfloat[]{%
        \includegraphics[width=0.8\linewidth]{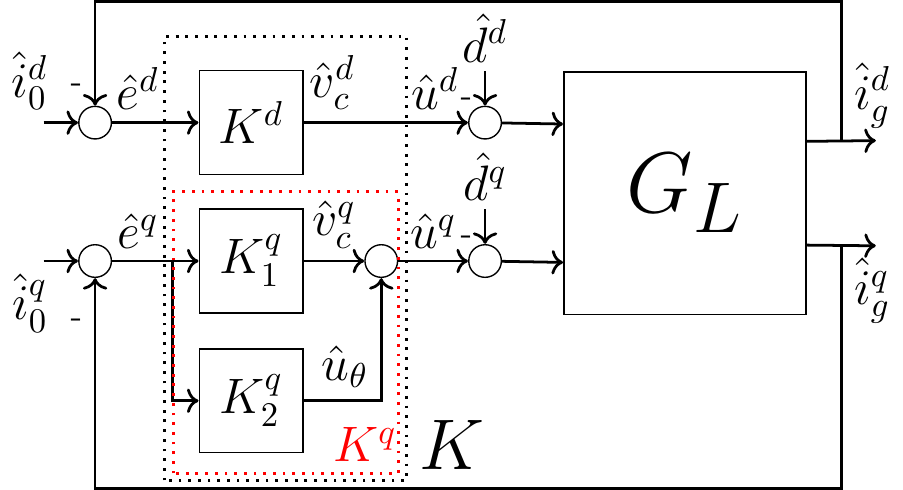}
        \label{fig:q_control_split}}
  \caption{(a) The control structure with diagonal feedback controller $[K^d,K^q]$, MIMO line dynamics $G_L$ and exogenous inputs $[\hat{i}_0^d,\hat{i}_0^q]^{\top}$, $[\hat{d}^d,\hat{d}^q]^{\top}$.}
  \label{fig:Feedback_Controller} 
  \vspace{-5mm}
\end{figure}
\section{A New Perspective on GFL Closed-Loop - MIMO Closed-Loop as a Perturbed 2-SISO System}
\label{sec:GFL_Closed-Loop_Dynamics_-_A_New Perspective}
In this work, we evaluate the stability, performance, transient response, and control design of the GFL inverter in terms of the closed-loop system that is formed around the line dynamics $G_L$ as shown in Fig. \ref{fig:q_control_split}. Here, $G_L$ is the MIMO transfer function in (\ref{eq:GL_expression}), and $K^d$ and $K^q=K_1^q+K_2^q$ form the diagonal feedback compensator $K=\text{diag}(K^d,K^q)$. Moreover, $\vv{e}=\vv{i_0} - \vv{i_g}$ denotes the tracking error, and $\vv{u}$ and $\vv{d}$ are the control effort and the input disturbance to the plant $G_L$. In the following proposition, we define the control effort $\vv{u} = [u^d,u^q]^{\top}$ and the input disturbance $\vv{d}=[d^d,d^q]^{\top}$ as they appear in Fig. \ref{fig:q_control_split} in terms of the capacitor voltage $\vv{v_c}$, $dq$ rotating frequency $\Dot{\theta}$, grid voltage $\vv{v_g}$, and grid frequency $\Dot{\theta}_g$. 
\begin{proposition}
\label{sec:disturbance}
For the closed-loop structure in Fig. \ref{fig:q_control_split}, the control effort $\vv{u}$ and input disturbance $\vv{d}$, are 
{\small\begin{align}
    \begin{bmatrix}
    u^d \\ u^q 
    \end{bmatrix}
     &=
    \begin{bmatrix}
    v_c^d \\ v_c^q
    \end{bmatrix} + 
    \begin{bmatrix}
    0 \\ u_\theta
    \end{bmatrix},
    &
    \begin{bmatrix}
    d^d \\ d^q
    \end{bmatrix} &= \|\vv{v_g}\|
    \begin{bmatrix}
    1 \\
    \int \Delta\omega_g \;dt
    \end{bmatrix},
    \label{eq:control_disturbance}
\end{align}}%
where
{\small\begin{align}
    u_\theta\left(t\right) &= v_0
    \int \Delta\omega \;dt,
    \label{eq:theta_definition}
    &
    \Delta\omega &= \Dot{\theta} - \omega_0,\;\;
    \Delta\omega_g = \Dot{\theta}_g - \omega_0,
\end{align}}%
and $\{v_0, \omega_0\}$ denote the nominal grid voltage (e.g. 120 V rms) and nominal grid frequency (e.g. 60 Hz).
\end{proposition}
\begin{remark}
In our proposed framework, we steer away from the conventional approach of using a separate PLL system, such as the one shown in Figs. \ref{fig:PLL_Conventional}, to directly control the $dq$ rotating angle $\theta$. Instead, $\theta$ is incorporated into the $K^q$ output signal $u^q = v_c^q + u_\theta$ as outlined in (\ref{eq:control_disturbance}) and (\ref{eq:theta_definition}), and shown in Fig. \ref{fig:q_control_split}. Our work takes advantage of three key benefits of this approach. First, our method enables us to formulate the GFL synchronization and frequency transient response as disturbance rejection objectives for the feedback controller (see Section \ref{sec: Conditions on SISO Sensitivity for Reference Tracking and Synchronization}). This perspective successfully eliminates the need for a separate PLL system. Second, by treating synchronization as a performance objective for the feedback controller, we are able to design the GFL closed-loop to achieve a desired balance between the stability margin and the quality of synchronization (see Section \ref{sec:water_bed}). Finally, our approach guarantees that a robustly stabilizing controller that meets the synchronization criteria will operate reliably in weak grid environments. It is important to note that when using a separate PLL, a robust controller alone does not ensure robust stability or even stability. The reason is that the interaction between the PLL and the grid can reduce the stability margin, leading to instability.
\end{remark}
\begin{remark}
The influence of the grid voltage $\vv{v}_g$ and frequency $\Dot{\theta}_g$ on the GFL closed-loop is treated as an input disturbance, denoted by $\vv{d}$ in (\ref{eq:control_disturbance}), to the plant as shown Fig. \ref{fig:q_control_split}. Considering the decentralized approach to GFL operation in this study, we assume that the magnitude of the grid voltage $\|\vv{v}_g\|$ and the grid frequency $\Dot{\theta}_g$, in (\ref{eq:control_disturbance}) and (\ref{eq:theta_definition}), are not known or measured. This assumption precludes the use of feedforward methods for mitigating grid disturbances. However, by implementing a disturbance rejection control framework, we can significantly mitigate the effects of disturbance $\vv{d}$. As a result, it is practical to base our control design in this paper on a nominal disturbance model. We proceed with the assumption that $\|v_g\|$ and $\Dot{\theta}_g$ are generally constant, but may experience deviations, harmonics, and other disturbances common in weak grid scenarios. Therefore, in line with (\ref{eq:control_disturbance}), we model the input disturbance as 
{\small\begin{equation}
    \begin{split}
        \begin{bmatrix}
            \hat{d}^d \\ \hat{d}^q
        \end{bmatrix} \approx v_0
        \begin{bmatrix}
            1/s \\
            \Delta\omega_g/s^2
        \end{bmatrix}
        + \vv{\eta}(s),
    \end{split}
    \label{eq:disturbance_type}
\end{equation}}%
where $\hat{d}^d$ is a constant and $\hat{d}^q$ is a ramp disturbance. Additionally, $\eta(s)$ represent the unmodeled disturbances. By adopting the nominal disturbance model in (\ref{eq:disturbance_type}), we prioritize the attenuation of disturbances within a specific frequency domain profile, essentially tailoring the performance characteristics of the controller to the expected disturbance.
\end{remark}
In the context of Figure \ref{fig:q_control_split}, the closed-loop transfer functions between the exogenous inputs $\{\vv{i_0},\vv{d}\}$ and the grid-current $\vv{i_g}$, tracking error $\vv{e}$, and control effort $\vv{u}$ are
{\small\begin{align}
    \vv{\hat{i}_g}&= \vv{\hat{i}_0} -  
    S(\vv{\hat{i}_0} + G_L \vv{\hat{d}}),
    & \vv{\hat{e}}&= 
    S(\vv{\hat{i}_0} + G_L \vv{\hat{d}}),
    \label{eq:Tracking_Error}\\
            \vv{\hat{u}} &= K 
    S(\vv{\hat{i}_0} + G_L \vv{\hat{d}}),
    \label{eq:Control_Effort}
\end{align}}%
where
{\small\begin{align}
    S &= (I_2 + G_L K)^{\text{-}1},
    \label{eq:Sensitivity_MIMO}
\end{align}}%
is the closed-loop MIMO {\em sensitivity} transfer function and $I_2$ denotes the $2\times 2$ identity matrix.\\
For a given feedback controller $K$, (\ref{eq:Tracking_Error}), (\ref{eq:Control_Effort}), and (\ref{eq:Sensitivity_MIMO}) uniquely determine the closed-loop stability, steady-state performance, and transient response to grid disturbance $\vv{d}$. However, the MIMO characteristics of $G_L$ and $S$ make the control design and analysis of closed-loop systems in (\ref{eq:Tracking_Error}) and (\ref{eq:Control_Effort}) significantly more intricate than a SISO system. Therefore, optimization-based methods, such as $\mathcal{H}_\infty$, are well suited to address MIMO control synthesis. However, in this paper, we use innovative algebraic manipulations to isolate the effect of coupled MIMO dynamics on closed-loop stability and performance. We achieve this by transforming the MIMO system into a nominal 2-SISO system with a multiplicative perturbation that embodies the effect of coupled MIMO dynamics. As we show throughout this paper, the proposed method enables us to address the performance and stability of MIMO closed-loop in terms of a nominal 2-SISO system.\\
The feedback controller $K$ in Fig. \ref{fig:q_control_split} is diagonal by design choice, and the MIMO nature of closed-loop is due to $G_L$ in (\ref{eq:GL_expression}). We decompose $G_L$ into a SISO part $\widetilde{G}_L$, and multiplicative coupled perturbation $(I_2 + E)$ as 
{\small  \begin{align}
    G_L &= \widetilde{G}_L(I_2 + E), 
    \label{eq:GL_factorization}
\end{align}}%
where
{\small\begin{align}
    \widetilde{G}_{L} &=
    \frac{\left(s + \lambda\right)/L}
    {s^2 + 2\lambda s + \lambda^2 + \omega_0^2},
    &
    E &= ( G_{L} - \widetilde{G}_{L} I_2
    )\widetilde{G}_{L}^{\text{-}1}.
    \label{eq:GL_tilde}
\end{align}}%
Using the SISO factorization of plant in (\ref{eq:GL_factorization}), we break down the closed-loop sensitivity in (\ref{eq:Sensitivity_MIMO}) into a 2-SISO component and a coupled MIMO perturbation.
\begin{proposition}
\label{prop: decoupling}
(a) \emph{SISO Factorization of Sensitivity:} We factor the MIMO sensitivity transfer function $S$ in (\ref{eq:Sensitivity_MIMO}) into a 2-SISO sensitivity transfer function $\widetilde{S}$ and coupled MIMO dynamics $\mathcal{X}_c$ and $\Gamma$ as
{\small\begin{align}
    S &= \widetilde{S}\mathcal{X}_{c}\Gamma,
    \label{eq:sensitivity_gamma}
\end{align}}%
where the 2-SISO sensitivity transfer function $\widetilde{S}$ is
{\small\begin{align}
    \widetilde{S} &= 
    \begin{bmatrix}
    \widetilde{S}^d & 0\\
    0 & \widetilde{S}^q
    \end{bmatrix} =
    \begin{bmatrix}
    (1 + \widetilde{G}_L K^d)^{\text{-}1} & 0 \\
    0 & (1 + \widetilde{G}_L K^q)^{\text{-}1}
    \end{bmatrix},
    \label{eq:SISO_Sensitivity}
\end{align}}%
and MIMO dynamics $\mathcal{X}_c$ and $\Gamma$ are
{\small\begin{align}
    \mathcal{X}_c &= (I_2 + (\Gamma - I_2)\widetilde{S}
    )^{\text{-}1},
    \label{eq: cross_mimo_dynamic}
    \\
    \Gamma &=
    \widetilde{G}_{L}G_{L}^{\text{-}1}= I_{2} +
    \frac{
    \begin{bmatrix}
        -\omega_{0}^{2} & 
        -\omega_{0}\left(s+\lambda\right)\\
        \omega_{0}\left(s+\lambda\right) & 
        -\omega_{0}^{2}
    \end{bmatrix}
    }{s^2+2\lambda s+ \lambda^2 + \omega_{0}^{2}}.
    \label{eq: PRGA}
\end{align}}
(b) \emph{SISO Characterization of Closed-loop Dynamics:}  The effect of the 2-SISO sensitivity $\widetilde{S}$ and coupled MIMO dynamics $\{\mathcal{X}_c,\Gamma\}$ on the GFL closed-loop dynamics in (\ref{eq:Tracking_Error}) and (\ref{eq:Control_Effort}) is explicitly given as
{\small\begin{align}
    \vv{\hat{i}_g} &= 
    \vv{i_0} - 
    \mathcal{X}_c\widetilde{S}
    \left(\Gamma\vv{\hat{i}_0} + \widetilde{G}_L\vv{\hat{d}}\right),\;
    \vv{\hat{e}} = 
    \mathcal{X}_c\widetilde{S}
    \left(\Gamma\vv{\hat{i}_0} + \widetilde{G}_L\vv{\hat{d}}\right),
    \label{eq:Tracking_Error_Diag}
    \\
    \vv{\hat{u}} &= K
    \mathcal{X}_c\widetilde{S}
    \left(\Gamma\vv{\hat{i}_0} + \widetilde{G}_L\vv{\hat{d}}\right).
    \label{eq:Control_Effort_Diag}
\end{align}}%
\\
(c) \emph{Bounding the Closed-loop Coupling:} We can use the SISO sensitivity $\widetilde{S}$ in (\ref{eq:SISO_Sensitivity}) to bound and reduce the effect of the MIMO coupled dynamic $\mathcal{X}_c$ on the closed-loop dynamics of the GFL in (\ref{eq:Tracking_Error_Diag}) and (\ref{eq:Control_Effort_Diag}). In more precise terms, we have
{\small\begin{align}
    \forall \widetilde{S}(j\omega)
    &\; \text{s.t.} \;
    \epsilon(\omega):=
    \|\widetilde{S}\left(j\omega\right)\|_2 
    \|\Gamma\left(j\omega\right) - I_2\|_{2}
    < 1,
    \label{eq:magnitude_condition}
    \\
    &\implies
    \|\mathcal{X}_c(j\omega) - I_2 \|_2
    \leq
    \frac{\epsilon\left(\omega\right)}{1 - \epsilon\left(\omega\right)}.
    \label{eq:perturbation_upper_bound}
\end{align}}%
\end{proposition}
\begin{remark}
Proposition \ref{prop: decoupling} touches on three crucial objectives. First, it deconstructs the MIMO sensitivity $S$ into a SISO and coupled dynamics. Second, it details the influence of each component on closed-loop dynamics in (\ref{eq:Tracking_Error_Diag}) and (\ref{eq:Control_Effort_Diag}). Most importantly, it shows the method to mitigate the coupling effect by leveraging $\widetilde{S}$. Essentially, the bound established in (\ref{eq:perturbation_upper_bound}) leads to the subsequent decoupling condition
{\small\begin{align}
    \|\epsilon\|_\infty \ll 1
    \implies
    \|\mathcal{X}_c - I_2\|_\infty \ll 1.
    \label{eq:decoupling_epsilon}
\end{align}}%
Based on above, $\|\epsilon\|_\infty \ll 1$ results in $\mathcal{X}_c \approx I_2$, approximately decoupling (\ref{eq:Tracking_Error_Diag}) and (\ref{eq:Control_Effort_Diag}) into
\begin{align}
    \vv{\hat{e}} &= \widetilde{S}
    (\Gamma\vv{\hat{i}_0} + 
    \widetilde{G}_L\vv{\hat{d}}),
    &
    \vv{\hat{u}} &= K\widetilde{S}
    (\Gamma\vv{\hat{i}_0} + 
    \widetilde{G}_L\vv{\hat{d}}).
    \label{eq: SISO_Closed_Loop}
\end{align}%
\end{remark}
\begin{remark}
Referring to the definition of $\epsilon$ in (\ref{eq:magnitude_condition}), the decoupling condition in (\ref{eq:decoupling_epsilon}) is equivalent to the following upper bound on the magnitude of the 2-SISO sensitivity 
{\small
\begin{equation}
    \begin{split}
        \|\widetilde{S}\left(j\omega\right)\|_2 <
        \frac{1}
        {\|\Gamma(j\omega) - I_2\|_2} =
        \frac{\left\|\left(\omega + j\lambda\right)^2 - \omega_0^2\right\|_2}
        {\omega_0
        \sqrt{\left(\omega + \omega_0\right)^2 + \lambda^2}},\; 
        \forall \omega.
    \end{split}
    \label{eq:coupling_stability_upper_bound}
\end{equation}}%
The above bound is uniquely shaped by $\lambda$ and, as shown in Fig. \ref{fig:lambda_stability}, the bound becomes less stringent (allowing larger magnitudes of $\widetilde{S}$) when $\lambda$ is higher, which corresponds to predominantly resistive line impedances. This relaxation also occurs as $\omega$ increases beyond the fundamental frequency $\omega_0$, regardless of the line characteristics $\lambda$. Essentially, to meet the decoupling condition, it is sufficient to maintain a small sensitivity within the lower frequency range, typically $[0,2\omega_0]$. Note that reducing the sensitivity $\widetilde{S}(j\omega)$ at any frequency is equivalent to raising the controller gain at the same frequency, as evident in (\ref{eq:SISO_Sensitivity}).
\end{remark}
In the rest of this work, we use the SISO sensitivity transfer function $\widetilde{S}$ in (\ref{eq:SISO_Sensitivity}) and the decoupled closed-loop dynamics in (\ref{eq: SISO_Closed_Loop}) as a basis for stability and performance analysis.
\vspace{-3mm}
\section{Characterization of Stability, Performance, and Fundamental Limitations on Control Design in terms of Nominal SISO Sensitivity}
\label{sec:Characterization_of_Stability_Performance}
\subsection{Conditions on SISO Sensitivity for Nominal Stability of GFL MIMO Closed-loop}
In this section we transform the nominal stability condition for the GFL's MIMO closed-loop into a set of simpler-to-check conditions based on the nominal SISO sensitivity $\widetilde{S}$ in (\ref{eq:SISO_Sensitivity}).
\begin{proposition}
\label{prop:stability}
The internal stability of the MIMO closed-loop in Fig. \ref{fig:q_control_split} is guaranteed if the following conditions hold.\\
(a) \emph{Stability of $\widetilde{S}$:} The two independent SISO sensitivities $\{\widetilde{S}^d,\widetilde{S}^q\}$ in (\ref{eq:SISO_Sensitivity}) have stable poles;\\
(b) \emph{Magnitude Condition:} $\epsilon$ in (\ref{eq:magnitude_condition})
satisfies the decoupling condition
$\|\epsilon\|_\infty< 1$, or equivalently, $\widetilde{S}$ satisfies (\ref{eq:coupling_stability_upper_bound}).
\end{proposition}
In the next section, we derive the robust stability conditions for the MIMO closed-loop based on the SISO sensitivity.%
\subsection{Conditions on SISO Sensitivity for Robust Stability of the GFL MIMO closed-loop}
\label{sec: SISO_Robust_Stability}
Proposition \ref{prop:stability} outlines the stability conditions for the GFL closed-loop based on the nominal value of the line impedance. However, it is often challenging to accurately measure or estimate the line impedance. Furthermore, in the case of a weak grid, the line impedance is prone to change with the grid condition. Our control framework provides robust stability (RS) even in the face of uncertain line impedance. It functions effectively under weak grid conditions and remains stable without the need for precise knowledge of the line impedance.\\
Typically, the exact value of the line impedance is not known, but we can specify the uncertainty intervals $\Pi_L$ and $\Pi_R$ for the line inductance and the line resistance as
{\small\begin{align}
    L \in
    \Pi_L = [L_{min},L_{max}],
    \quad
    R \in
    \Pi_R = [R_{min},R_{max}],
    \label{eq:uncertainty_interval}
\end{align}}%
where the minimum and maximum anticipated line inductance and resistance are denoted by $\{L_{min},L_{max}\}$, and $\{R_{min},R_{max}\}$, respectively. In this context, a robustly stabilizing feedback controller must satisfy Proposition \ref{prop:stability} for all possible SISO plants $\widetilde{G}_L$ that are generated by sets $\Pi_L$ and $\Pi_R$ in (\ref{eq:uncertainty_interval}). We achieve this by forming the set that encompasses all possible SISO plants, which we label $\Pi_G$. Subsequently, we define the nominal plant, denoted $\widetilde{G}_{L0}$, for the impedance uncertainty intervals outlined in (\ref{eq:uncertainty_interval}). Finally, we detail the RS conditions for the set $\Pi_G$ based on the nominal SISO sensitivity formed by the nominal plant $\widetilde{G}_{L0}$.
\\
The set of all possible SISO line dynamics, denoted as $\Pi_G$, is determined by the uncertainty sets $\Pi_R$ and $\Pi_L$ in (\ref{eq:uncertainty_interval}) as
{\small\begin{align}
    \Pi_G = 
    \left\{
    \widetilde{G}_L(s) = \frac{(s + \lambda)/L}
    {s^2 + 2\lambda s + \lambda^2 + \omega_2} \,\bigg|\,
    L \in \Pi_L, \lambda \in \Pi_\lambda
    \right\},
    \label{eq:Plant_Uncertain_Set}
\end{align}}%
where
{\small\begin{align}
    \Pi_\lambda = 
    \left[\lambda_{min}= \frac{R_{min}}{L_{max}},
    \lambda_{max}= \frac{R_{max}}{L_{min}}\right].
    \label{eq:uncertainty_lambda}
\end{align}}%
Moreover, we define the nominal SISO plant, represented by $\widetilde{G}_{L0}$, for the given uncertainty sets $\Pi_L$ and $\Pi_\lambda$ as
{\small\begin{align}
    \widetilde{G}_{L0} = 
    \frac{(s + \lambda_0)/L_0}
    {s^2 + 2\lambda_0 s + \lambda_0^2 + \omega_0^2},
    \label{eq:nominal_RS_Plant}
\end{align}}%
where
{\small\begin{align}
    L_0 &= 2\frac{L_{min} L_{max}}
    {L_{min} + L_{max}}, 
    &
    \lambda_0 = \frac{\lambda_{max}L_{max} + 
    \lambda_{min}L_{min}}
    {L_{min} + L_{max}}.
    \label{eq:nominal_RS_parameters}
\end{align}}%
Subsequently, the nominal SISO sensitivity for $\widetilde{G}_{L0}$ is
{\small\begin{align}
    \widetilde{S}_0 &= 
    \begin{bmatrix}
    \widetilde{S}_0^d & 0\\
    0 & \widetilde{S}_0^q
    \end{bmatrix} =
    \begin{bmatrix}
    (1 + \widetilde{G}_{L0} K^d)^{\text{-}1} & 0 \\
    0 & (1 + \widetilde{G}_{L0} K^q)^{\text{-}1}
    \end{bmatrix}.
    \label{eq:nominal_plant_sensitivity}
\end{align}}%
The feedback controller $K=\text{diag}(K^d,K^q)$ in (\ref{eq:nominal_plant_sensitivity}) is robustly stabilizing if the closed-loop system maintains stability despite any perturbation of the nominal plant $\widetilde{G}_{L0}$ in (\ref{eq:nominal_RS_Plant}) within the set of all possible perturbed line dynamics $\Pi_G$. In the following proposition, we present the robust stability conditions for the GFL MIMO closed-loop in terms of the nominal SISO sensitivity $\widetilde{S}_{0}$ in (\ref{eq:nominal_plant_sensitivity}).
\begin{proposition}
\label{prop: robust_stability}
The closed-loop in Fig. \ref{fig:q_control_split} is robustly stable (RS) with respect to the set of all possible line dynamics $\Pi_G$ in (\ref{eq:Plant_Uncertain_Set}) if the nominal sensitivity $\widetilde{S}_0$ in (\ref{eq:nominal_plant_sensitivity}) satisfies: \\
(a) the stability conditions in part (a) of Proposition \ref{prop:stability},
\\
(b) and the following two bounds
{\small\begin{align}
    \|
    \widetilde{S}_0(j\omega)
    \|_2 
    &< 
    \frac{1 - \|W_1(j\omega)\|_2 \|W_3(j\omega)\|_2}
    {\|W_2(j\omega)\|_2}, 
    \quad \forall\omega,
    \label{eq:first_RS_condition}
    \\
    \|
    \widetilde{S}_0(j\omega)
    \|_2 
    &<
    \frac{\left\|\left(\omega + j\lambda_{min}\right)^2 - \omega_0^2\right\|_2}
    {\omega_0
    \sqrt{\left(\omega + \omega_0\right)^2 + \lambda_{min}^2}},\; 
    \quad \forall \omega.
    \label{eq:second_RS_condition}
\end{align}}%
where
{\small\begin{align}
    W_1 &= 
    \frac{(1/L_0 - 1/L_{max})s + 
    (\lambda_0/L_0 - \lambda_{min}/L_{max})}
    {(s + \lambda_0)/L_0},
    \label{eq:w1_design}
    \\
    W_2 &= 
    \frac{2(\lambda_{max} - \lambda_0)s +
    (\lambda_{max}^2 - \lambda_0^2)}
    {s^2 + 2\lambda_0 s + 
    \lambda_0^2 + \omega_0^2},
    \; \text{and} \;
    W_3 = \frac{\omega_{bw}}{s + \omega_{bw}}.
    \label{eq:w2_design}
\end{align}}%
In this proposition, $\{L_0,\lambda_0\}$ are the nominal parameters of (\ref{eq:nominal_RS_parameters}), $\{L_{max},\lambda_{max},\lambda_{min}\}$ are the same as in (\ref{eq:uncertainty_interval}) and (\ref{eq:uncertainty_lambda}), and $\omega_{bw}$ represents the closed-loop bandwidth for the nominal SISO system.
\end{proposition}
\begin{figure}
    \centering
    \subfloat[]{
    \includegraphics[width=0.495\linewidth]{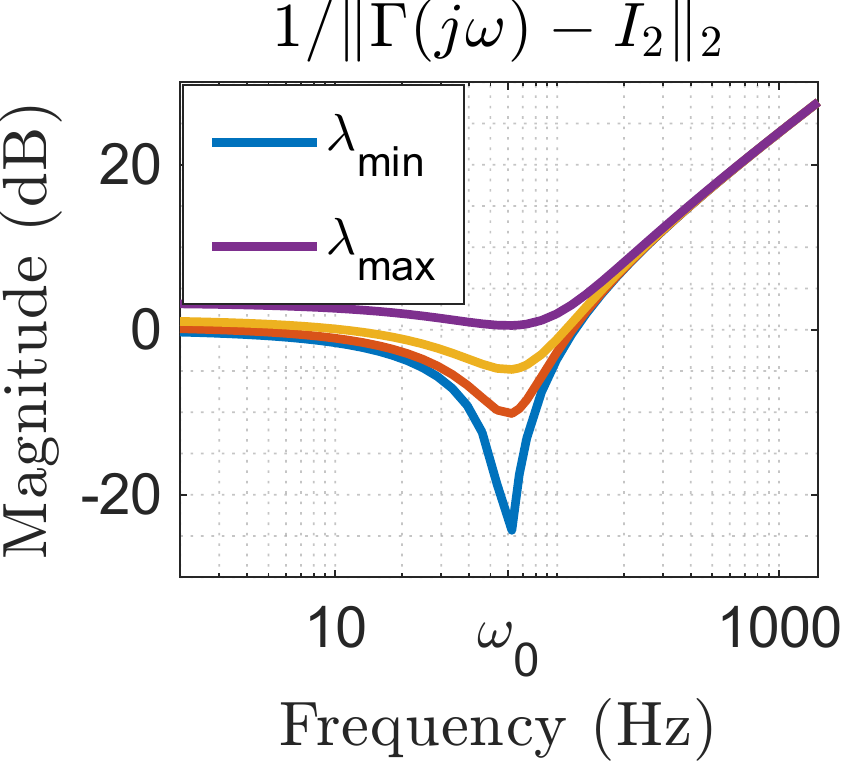}
    \label{fig:lambda_stability}
    }
    \subfloat[]{
    \includegraphics[width=0.48\linewidth]{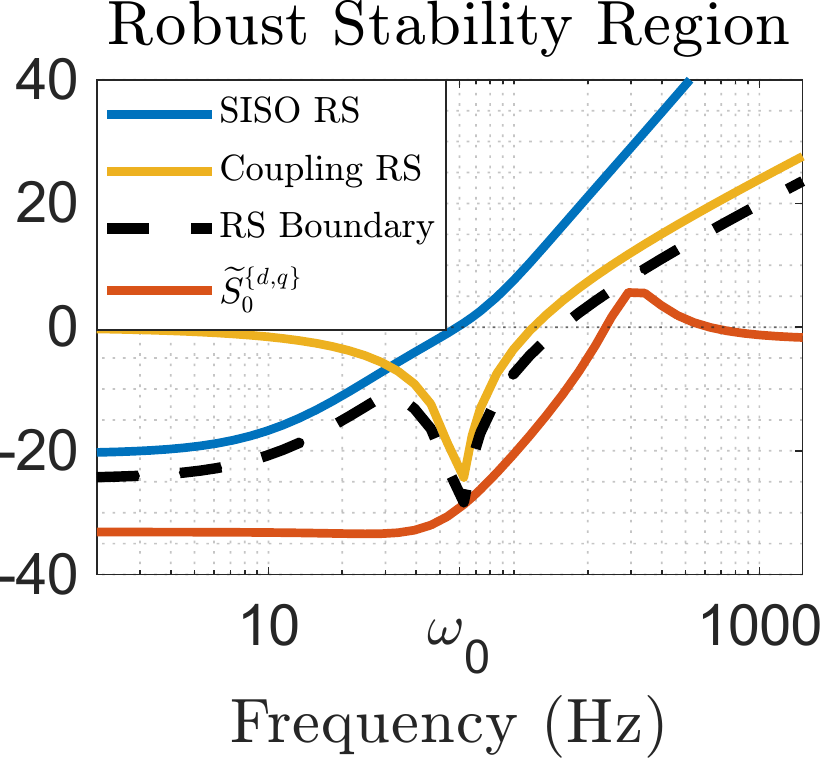}
    \label{fig:RS_Region}
    }
    \caption{(a) The upper bound in (\ref{eq:coupling_stability_upper_bound}) for different values of $\lambda$. The lower values of $\lambda$ correspond to smaller upper bounds. (b) The RS boundary, as shown by the dashed line, consists of two distinct upper bounds. The SISO upper bound (\ref{eq:first_RS_condition}) and coupling upper bound. Any nominal closed-loop sensitivity $\widetilde{S}_0$ that remains below the RS boundary is considered to have RS property.}
    \vspace{-3mm}
\end{figure}%
\begin{remark}
The upper bound outlined in (\ref{eq:first_RS_condition}) is based on the RS condition for the nominal SISO closed-loop. On the other hand, (\ref{eq:second_RS_condition}) arises from the RS condition for the perturbations of the coupling dynamics, as detailed in the proof. As illustrated in Fig. \ref{fig:RS_Region}, the RS bound for the SISO closed-loop tends to be more restrictive at lower frequencies. Meanwhile, the RS constraint for coupling dynamics becomes more significant around and above the fundamental frequency $\omega_0$. Furthermore, when line inductance increases (which corresponds to a smaller $\lambda$ value), the constraint in (\ref{eq:coupling_stability_upper_bound}) becomes more restrictive at higher frequencies. This trend is clearly depicted in Fig. \ref{fig:lambda_stability}. As a result of these observations, the RS conditions outlined in this section indicate that, particularly in inductive lines, the closed-loop system should maintain a minimum bandwidth of $2\omega_0$ ($120$ Hz) to ensure robust stability.
\end{remark}%
\subsection{Conditions on SISO Sensitivity for Reference Tracking and Synchronization}
\label{sec: Conditions on SISO Sensitivity for Reference Tracking and Synchronization}
The primary objective of the GFL inverter is to \emph{synchronize} and \emph{inject} the required power into the grid while maintaining \emph{power quality} (i.e. suppress harmonic and high-frequency disturbances). These objectives are given concisely in terms of closed-loop in Fig. \ref{fig:q_control_split} as
{\small\begin{align}
    \lim_{t \to \infty} \vv{e}(t) &= 0,
    &
    \lim_{t \to \infty} v_c^q(t) &= 0,
    &
    \frac{\|\vv{\hat{e}}(j\omega)\|_2}
    {\|\vv{\hat{w}}(j\omega)\|_2}
    &\ll 1, \;
    \forall \omega,
    \label{eq: GFL_Objectives}
\end{align}}%
where $\vv{\hat{w}} = \Gamma\vv{\hat{i}_0} + \widetilde{G}_L\vv{\hat{d}}$. In the above, $\vv{e} = 0$ represents the desired zero steady-state tracking error, and $v_c^q =0$ signifies the steady-state synchronization that decouples the mapping between power and current in (\ref{eq:power_current_mapping}). Furthermore, the inequality in (\ref{eq: GFL_Objectives}) represents a key objective aimed at reducing the effect of closed-loop input signals, represented by $\vv{\hat{w}}$, on the tracking error $\vv{\hat{e}}$ throughout the frequency spectrum. This is crucial to ensure that the current waveform remains clear and free of high-frequency distortions.\\
The GFL objectives in (\ref{eq: GFL_Objectives}) can be formulated in terms of closed-loop SISO sensitivity transfer function $\widetilde{S}$ as
{\small\begin{align}
    \lim_{s \to 0} \widetilde{S}
    \vv{\hat{w}}s &= 0,
    &
    \lim_{s \to 0} [0\; K_1^q]\widetilde{S}
    \vv{\hat{w}}s
    &= 0,
    &
    \|\widetilde{S}(j\omega)
    \|_2 &\ll 1, \;
    \forall \omega.
    \label{eq: GFL_Objectives_Laplace}
\end{align}}%
The above equation is based on applying the final value theorem to $\vv{\hat{e}}$ in (\ref{eq: SISO_Closed_Loop}), and $\hat{v}_c^q = [0\;K_1^q]\vv{e}$ (see Fig. \ref{fig:q_control_split}). The inequality in (\ref{eq: GFL_Objectives_Laplace}) simply follows from (\ref{eq: SISO_Closed_Loop}). The following proposition uses the definition of $\widetilde{S}$ in (\ref{eq:SISO_Sensitivity}) and the grid disturbance model in (\ref{eq:disturbance_type}) to convert the objectives in (\ref{eq: GFL_Objectives_Laplace}) into the conditions of the feedback controller $K$.
\begin{proposition}
\label{prop:synchronization}
(a) We achieve $\lim_{t \to \infty}\vv{e}=0$, if and only if $\widetilde{S}^d$ has at least one zero and $\widetilde{S}^q$ has at least two zeros at the origin. This is equivalent to $K^d$ having at least one pole and $K^q$ having at least two poles at the origin. \\
(b) We achieve $\lim_{t  \to \infty}v_c^q=0$ (synchronization) if and only if $K_1^q/K_2^q$ in Fig. \ref{fig:q_control_split} possesses at least two zeros at the origin.\\
(c) We can reduce the effect of grid side disturbance $\vv{d}$ on the tracking error $\vv{e}$ at any specific frequency, using a high-gain feedback controller. This approach is effective within the closed-loop bandwidth, and the feedback controller $K$ imposes the following upper bound on the attenuation factor
{\small\begin{align}
    \frac{\|\vv{\hat{e}}(j\omega)\|_2}
    {\|\vv{\hat{d}}(j\omega)\|_2} 
    \leq
    \frac{1}
    {\|K(j\omega)\|_2},
    \quad \forall \omega.
    \label{eq: Controller_Disturbance_Attenuate}
\end{align}}%
\end{proposition}
\begin{remark}
The above proposition formulates the synchronization objective as a disturbance rejection characteristic of the feedback controller. It also outlines the necessary and sufficient conditions that the controller must satisfy to achieve synchronization. This eliminates the need for separate PLL and allows for a unified closed-loop analysis between the synchronization quality, stability, and performance objectives.
\end{remark}
\begin{remark}
    Propositions \ref{prop:stability}, \ref{prop: robust_stability}, and \ref{prop:synchronization} suggest that a high-gain controller can ensure robust stability and produce a clean and distortion-free current waveform by reducing the sensitivity $\widetilde{S}$ across the frequency spectrum. However, in practice, implementing a controller that maintains substantial gain across a broad frequency band is costly and introduces high-frequency noise into the system. Moreover, as discussed in Section \ref{sec:The_High_Frequency_Resonance}, a fundamental constraint called the Bode sensitivity integral (also known as the waterbed effect) limits the use of high-gain controller over a wide frequency range.
\end{remark}
\subsection{Conditions on Feedback Controller to Shape the Transient Response of Inverter Frequency}
\label{Controlled Frequency Transient Response}
DC/AC inverters do not possess inherent inertia; however, numerous studies, especially for grid-forming inverters, have mimicked the inertial response of synchronous generators. Our work provides a general framework that subsumes the inertial response of the inverter as part of the transient response of the closed-loop in Fig. \ref{fig:q_control_split} to the grid frequency disturbance $d^q$.
\begin{proposition}
\label{prop:freq_transient}
The transient and steady-state response of the inverter's frequency deviation $\Delta\omega$ to the grid frequency deviation $\Delta\omega_g$ is uniquely specified by
{\small\begin{align}
    \Delta \omega &=
    \widetilde{T}_\theta^q
    \Delta\omega_g,
    &
    \widetilde{T}_\theta^q &=
    K_2^q\widetilde{G}_L\widetilde{S}^q =
    \frac{K_2^q\widetilde{G}_L}
    {1 + \widetilde{G}_L(K_1^q + K_2^q)}.
    \label{eq: Inverter_Frequency_TF}
\end{align}}%
Moreover, $\widetilde{T}_\theta^q$ always assumes the shape of a low-pass filter with unity dc gain.
\end{proposition}
\begin{remark}
\label{remark:Frequency_Transient}
The step response of $\widetilde{T}_\theta^q$, such as the rise time, the settling time, the overshoot, and the oscillation, are generalizations of concepts such as the rate of change of frequency (RoCoF), the frequency nadir (or zenith), and the inertia. Therefore, we can shape the open loop $K_2\widetilde{G}_L$ in (\ref{eq: Inverter_Frequency_TF}) to obtain the desired transient response to grid frequency disturbance $\Delta\omega_g$. Specifically:
\\
$\bullet$ \textbf{Cross-over frequency:}
The cross-over frequency of $K_2^q\widetilde{G}_L$ affects both RoCoF and inertia. A lower cross-over frequency improves the inertial response, reduces RoCoF, and attenuates disturbances and harmonics on the inverter frequency.
\\
$\bullet$ \textbf{Phase margin:}
The phase margin of $K_2^q\widetilde{G}_L$ influences the overshoot, where a higher phase margin dampens the transient response and reduces the frequency nadir or zenith.
\\
$\bullet$ \textbf{Roll-off rate}
The roll-off rate of $K_2^q\widetilde{G}_L$ above the cross-over frequency specifies the attenuation factor of the grid side disturbances and harmonics on the inverter frequency.
\end{remark}
\vspace{-3mm}
\section{Fundamental Limitations on Control Design and Trade-Off Between Distinct Objectives}
\label{sec:water_bed}
We use the sensitivity $\widetilde{S}$ to illustrate two primary bottlenecks in the GFL closed-loop control. These limitations and trade-offs are fundamental and cannot be solved regardless of the control design. However, by understanding them, we can design controllers that can achieve the desired trade-off.
\subsection{The Disturbance Attenuation Bottleneck}
We cannot attenuate the impact of grid disturbance $\vv{d}$ on current tracking error $\vv{e}$, inverter voltage $\vv{v_c}$, and inverter frequency $\Dot{\theta}$ simultaneously. In fact, reducing the effect of disturbance on one variable inevitably amplifies the disturbance on the other two. We validate this by rewriting the transfer functions between $\vv{d}$ and $\{\vv{e},\vv{u}\}$ in (\ref{eq: SISO_Closed_Loop}) as
{\small\begin{align}
    \begin{bmatrix}
        \hat{e}^d \\ \hat{e}^q
    \end{bmatrix}
    &= \widetilde{G}_L
    \begin{bmatrix}
        \widetilde{S}^d & 0 
        \\
        0 & \widetilde{S}^q
    \end{bmatrix}
    \begin{bmatrix}
        \hat{d}^d \\ \hat{d}^q
    \end{bmatrix},
    &
    \begin{bmatrix}
        \hat{u}^d \\ \hat{u}^q
    \end{bmatrix}
    &=
    \begin{bmatrix}
        \widetilde{T}^d & 0 
        \\
        0 & \widetilde{T}^q
    \end{bmatrix}
    \begin{bmatrix}
        \hat{d}^d \\ \hat{d}^q
    \end{bmatrix}.
    \label{eq:disturbance_attenuation_bottleneck}
\end{align}}%
The transfer functions $\{\widetilde{T}^d, \widetilde{T}^q\}$ above are defined as 
{\small\begin{align}
    \widetilde{T}^d &=
    K^d\widetilde{G}_L\widetilde{S}^d,
    &
    \widetilde{T}^q
    &=
    K^q\widetilde{G}_L\widetilde{S}^q
    =
    (K_1^q + K_2^q)\widetilde{G}_L\widetilde{S}^q
    ,
    \label{eq:Td_Tq_TF}
\end{align}}%
and together with the sensitivities $\{\widetilde{S}^d,\widetilde{S}^q\}$, they satisfy the following algebraic constraints
{\small\begin{align}
    \widetilde{T}^d +
    \widetilde{S}^d
    &=
    1,
    &
    \widetilde{T}^q +
    \widetilde{S}^q
    &=
    1.
    \label{eq:current_control_effort_bottle_neck}
\end{align}}%
We want to minimize the magnitude of $\{\widetilde{S}^q,\widetilde{S}^q\}$ through a broad frequency range to reduce the impact of the disturbance $\vv{d}$ on the tracking error $\vv{e}$ in (\ref{eq:disturbance_attenuation_bottleneck}) and attain perfect current reference tracking (\ref{eq: SISO_Closed_Loop}). Furthermore, the decoupling and stability of the closed-loop MIMO dynamics in Propositions \ref{prop: decoupling} and \ref{prop:stability} are based on reducing the magnitude of $\{\widetilde{S}^q,\widetilde{S}^q\}$. However, based on the algebraic constraint in (\ref{eq:current_control_effort_bottle_neck}), reducing $\{\widetilde{S}^q,\widetilde{S}^q\}$ increases $\{\widetilde{T}^d,\widetilde{T}^q\}$ toward 1 creating a unity gain path between $\vv{d}$ and $\vv{u}$ in (\ref{eq:disturbance_attenuation_bottleneck}). Considering that $u^d=v_c^d$ and $u^q=v_c^q + u_\theta$, this demonstrates a fundamental trade-off in which attenuating the distortion and harmonics caused by grid disturbances $\vv{d}$ on the tracking error $\vv{e}$ amplifies the effect of grid disturbances on $u_\theta$ and $\vv{v_c}$.\\
In GFL operation, we are primarily concerned with the quality of the current waveform, so it is plausible to minimize $\{\widetilde{S}^d,\widetilde{S}^q\}$ over a wide frequency range and compromise the quality of the inverter voltage in favor of a harmonic and distortion-free current waveform. However, in the $q$ loop, this can lead to distortion of $u_\theta$ by high-frequency grid disturbances and potentially destabilize inverter operation. Our proposed closed-loop structure can solve this problem.
\subsubsection{Inverter Frequency Disturbance Immunity}
\label{remark:freq_volt_bottleneck}
\begin{figure}[t]
    \centering
    \subfloat[]{\includegraphics[width = 0.95\linewidth]{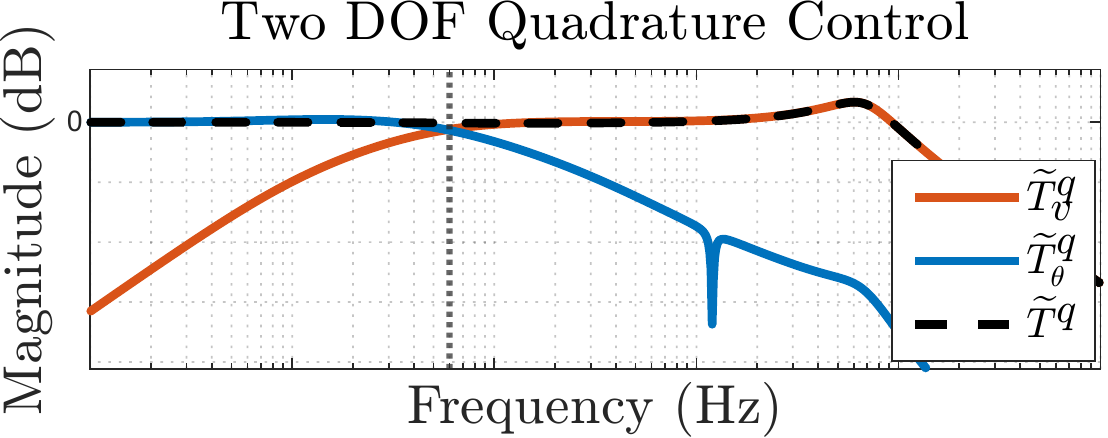}
    \label{fig:composit_response}}
    \caption{(a) The quadrature closed-loop $\widetilde{T}^q$ is comprised of low-pass $\widetilde{T}_{\theta}^q$ and band-pass $\widetilde{T}_{v}^q$. This attenuates disturbances on $u_\theta$ beyond $\widetilde{T}_{\theta}^q$ cut-off frequency (dotted line) by shifting the high-frequency disturbances to $v_c^q$.}
\end{figure}

The $dq$ signals are calculated based on the angle of the rotating frame. Therefore, any disturbance and harmonics in $\theta$ will distort the $dq$ signals and potentially destabilize the inverter. In this section, we leverage the parallel structure of the $q$ axis controller, shown in Fig. \ref{fig:q_control_split}, to enhance the resilience of $u_\theta$ and consequently $\theta$ against grid disturbances.\\
As discussed in Proposition \ref{prop:freq_transient}, $\widetilde{T}_\theta^q$ forms a unity-gain low-pass filter between $\Delta\omega_g$ and $\Delta\omega$. Therefore, it attenuates the impact of grid distortions and harmonics above the cutoff frequency. Subsequently, reducing the bandwidth of $\widetilde{T}_\theta^q$ improves the robustness of $\theta$ against high-frequency grid disturbances. Simultaneously, based on (\ref{eq: GFL_Objectives_Laplace}) and (\ref{eq:disturbance_attenuation_bottleneck}), we want to minimize $\widetilde{S}^q$ across a broad frequency range to achieve harmonic and distortion-free $\hat{e}^q$. Balancing these two requirements - low bandwidth for $\widetilde{T}_\theta^q$ and minimize $\widetilde{S}^q$ across a wide frequency range - imposes a unique design challenge for the parallel quadrature controller $K^q = K_1^q + K_2^q$. We formalize this design constraint on the $q$ axis controller by rewriting the algebraic constraint in (\ref{eq:current_control_effort_bottle_neck}) as
{\small\begin{align}
    1 - \widetilde{S}^q =
    \widetilde{T}^q = 
    \widetilde{T}_\theta^q + \widetilde{T}_v^q,
    \label{eq:Frequency_T_Bottleneck}
\end{align}}%
where 
{\small\begin{align}
    \hat{v}_c^q &= \widetilde{T}_v^q \hat{d}^q,
    &
    \widetilde{T}_v^q &= 
    K_1^q\widetilde{G}_L\widetilde{S}^q =
    \frac{K_1^q\widetilde{G}_L}
    {1 + \widetilde{G}_L(K_1^q + K_2^q)},
    \label{eq:d_to_vcq}
\end{align}}%
is the transfer function between $d^q$ and $v_c^q$. The algebraic constraint in (\ref{eq:Frequency_T_Bottleneck}) suggests that the sum of $\widetilde{T}_\theta^q$ and $\widetilde{T}_v^q$ should equal one over a broad frequency range. This range is defined by the bandwidth of the closed-loop transfer function $\widetilde{T}^q$. Therefore, $\widetilde{T}_v^q$ needs to function as a bandpass filter with unity gain between the cutoff frequencies of $\widetilde{T}_\theta^q$ and $\widetilde{T}^q$, as illustrated in Fig. \ref{fig:composit_response}. Taking into account the parallel structure of $K^q$ and the definition of $\widetilde{T}_v^q$ and $\widetilde{T}_\theta^q$, $K_1^q$ and $K_2^q$ should satisfy the following constraint within closed-loop bandwidth
{\small\begin{align}
    \widetilde{T}_\theta^q &\approx
    \frac{K_2^q}{K_1^q + K_2^q}
    \,\text{(Low-Pass)}
    &
    \widetilde{T}_v^q &\approx
    \frac{K_1^q}{K_1^q + K_2^q} 
    \,\text{(Band-Pass)}.
\end{align}}%
Intuitively, $K_2^q$ should have high-gain within the low frequency, while $K_1^q$ should have high-gain in the band-pass region.
\subsubsection{Operation Under Asymmetrical Fault}
\label{sec: asymmetrical_fault}
During an asymmetrical fault, the PCC voltage becomes unbalanced. Under these conditions, we can represent the unbalanced three-phase voltage as a superposition of the positive sequence, the negative sequence, and the zero sequence as \cite{yazdani2010voltage}
{\small\begin{alignat}{3}
    v_g^a &= a \|v_g\| \cos{(\Dot{\theta}_g t)}
    &&+ b\|v_g\| \cos{(\Dot{\theta}_g t + \psi)} 
    &&+ v_g^0,
    \nonumber
    \\
    v_g^b &= a \|v_g\| \cos{(\Dot{\theta}_g t - \frac{2\pi}{3})}
    &&+ b\|v_g\| \cos{(\Dot{\theta}_g t + \psi - \frac{4\pi}{3})} 
    &&+ v_g^0,
    \nonumber
    \\
    v_g^c &= a \|v_g\| \cos{(\Dot{\theta}_g t - \frac{4\pi}{3})}
    &&+ b\|v_g\| \cos{(\Dot{\theta}_g t + \psi - \frac{2\pi}{3})} 
    &&+ v_g^0,
    \label{eq:asymmetric_fault}
\end{alignat}}%
where $a$ and $b$ are amplitudes of positive- and negative-sequence unbalanced voltages with respect to the $\|v_g\|$. Furthermore, $v_g^0$ represents the zero sequence and $\psi$ is the phase angle of the negative sequence with respect to the positive sequence. In a synchronized $dq$ frame, (\ref{eq:asymmetric_fault}) is \cite{6739413}
{\small\begin{align}
    \begin{split}
        v_g^d &= \|v_g\| \left(
        a\cos{(\delta)} 
        + b \cos{(2\Dot{\theta}_g t - \delta + \psi)}
        \right),
        \\
        v_g^q &= \|v_g\| \left(
        a \sin{(\delta)} 
        - b \sin{(2\Dot{\theta}_g t - \delta + \psi)}
        \right),
    \end{split}
    \label{eq:asymmetric_dq}
\end{align}}%
where $\delta$ represent the phase angle between the grid positive-sequence voltage phasor and synchronized $dq$ frame as shown in Fig. \ref{fig:DQ_Frame}. Following along the lines of a Proposition \ref{sec:disturbance}, and based on (\ref{eq:asymmetric_dq}), the input disturbance under asymmetric fault is
{\small
\begin{align}
    \begin{bmatrix}
    d^d \\ d^q
    \end{bmatrix} =
    v_0 \left(
    a
    \begin{bmatrix}
    1 
    \\
    \int \Delta\omega_g\;dt
    \end{bmatrix} + 
    b
    \begin{bmatrix}
    \hphantom{-}\cos{\left(
    2\Dot{\theta}_g t - \delta +\psi \right)}
    \\
    -\sin{\left(
    2\Dot{\theta}_g t - \delta +\psi \right)}
    \end{bmatrix}\right).
\end{align}}%
Compared to the nominal disturbance model in (\ref{eq:control_disturbance}), the asymmetric fault imposes an additional second-harmonic component on the disturbance $\vv{d}$ in the $dq$ frame. The second harmonic, unchecked, propagates to the output current (\ref{eq: SISO_Closed_Loop}) and results in the third harmonic in the stationary frame \cite{yazdani2010voltage}. We can attenuate the effect of the second-harmonic disturbance on $\vv{\hat{e}}$ by reducing the sensitivity $\widetilde{S}$ at $2\Dot{\theta}_g$ (see (\ref{eq:SISO_Sensitivity}) and (\ref{eq:disturbance_attenuation_bottleneck})). As shown in (\ref{eq: Controller_Disturbance_Attenuate}), this is achieved by cascading a high-gain PR controller, as detailed below, with the controller $K$.
{\small\begin{align}
    K_{\text{PR}} &= 1 + k_2
    \frac{4\xi_2\omega_0 s}
    {s^2 + 4\xi_2\omega_0 s + 4\omega_0^2},
    &
    \xi_2 \geq 
    \max_{\Delta\omega_g} \left\|
    \frac{\Delta\omega_g}
    {\omega_0}\right\|.
    \label{eq:PR_Grid_Fault}
\end{align}}%
Based on (\ref{eq: Controller_Disturbance_Attenuate}), increasing the PR gain $k_2$ leads to a more effective attenuation of second-harmonic disturbances. Furthermore, the choice of the damping factor $\xi_2$ is influenced by the anticipated maximum deviation of the grid frequency from its nominal value $\omega_0$. When $\xi_2$ is set to a higher value, the PR controller becomes more efficient in mitigating harmonics caused by asymmetrical faults, especially in scenarios where the frequency of the grid strays from the nominal value, a common situation in weak grids. However, as we show in Section \ref{sec:The_High_Frequency_Resonance} and in experiments, higher values of $k_2$ and $\xi_2$ induce a high frequency resonance.\\
Cascading the PR compensator with the feedback controller $K$ is not sufficient to guarantee effective performance during an asymmetrical fault. Furthermore, it is necessary to maintain a harmonic-free rotational frame angle $\theta$. Based on the analysis in Section \ref{remark:freq_volt_bottleneck}, the algebraic constraint in (\ref{eq:Frequency_T_Bottleneck}) and the definition of $\widetilde{T}_v^q$ in (\ref{eq:d_to_vcq}), cascading the PR compensator with $K_1^q$ brings $\|\widetilde{T}_v^q(2j\omega_0)\|_2$ close to one, while $\widetilde{T}_\theta^q$ exhibits a notch behavior at $2\omega_0$ as shown in Fig. \ref{fig:composit_response}. The notch response of $\widetilde{T}_\theta^q$ at $2\omega_0$ allows the inverter frequency to filter out the second harmonic created by the asymmetric fault, allowing the fault ride through.
\begin{remark}
The decoupled double-synchronous reference frame-PLL (DDSRF-PLL) is an advanced algorithm that is used to accurately track the phase and frequency of the grid voltage under distorted or unbalanced conditions \cite{rodriguez2007decoupled,achlerkar2021new}. DDSRF-PLL employs two synchronous frames: one aligns with the positive-sequence and the other with the negative-sequence. This approach allows the DDSRF-PLL to effectively detect and synchronize with the grid frequency, particularly under unbalanced conditions. Our proposed control design and closed-loop structure can work under unbalanced grid conditions and has the same performance as DDSRF-PLL without the need for an extra synchronous reference frame. The robust stability argument in Section \ref{sec: SISO_Robust_Stability} guarantees the robustness of our approach in the face of grid uncertainties and fluctuations. In the experiments section, we compare DDSRF-PLL with our proposed approach in terms of performance and sensitivity to grid uncertainties.
\end{remark}
\subsection{The High-Frequency Resonance}
\label{sec:The_High_Frequency_Resonance}
\begin{figure}[t]
    \centering
    \subfloat[]{\includegraphics[width = 0.87\linewidth]{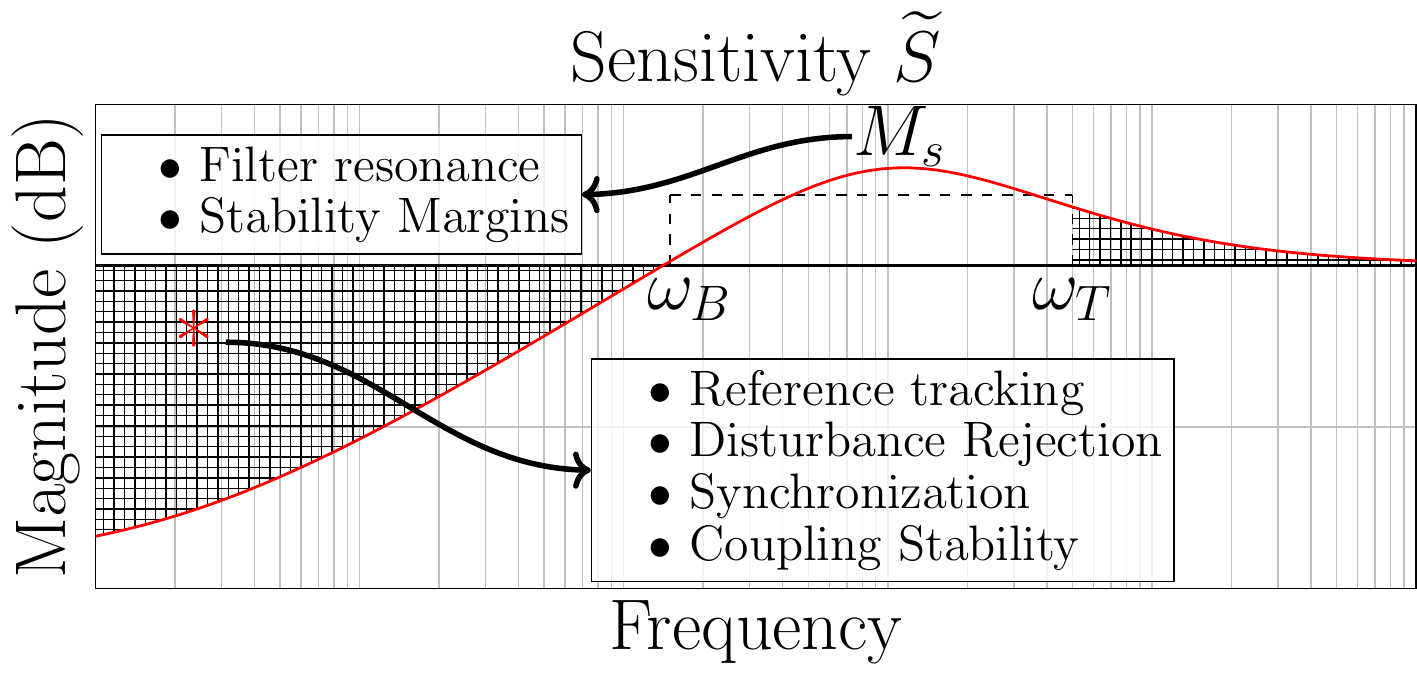}
    \label{fig:Water_Bed}}
    \caption{(a) The shape of $\widetilde{S}$ at different frequencies directly indicates performance objectives and stability margins.}
\end{figure}

Ideally, we want to keep the magnitude of sensitivity small at all frequencies to maintain perfect power tracking and disturbance rejection (\ref{eq:Tracking_Error}), minimizing the effect of coupling (\ref{eq:perturbation_upper_bound}) and robust stability (\ref{eq:coupling_stability_upper_bound}). However, the Bode sensitivity integral (\ref{eq:Bode_Sens}) indicates a fundamental limitation on the total amount of sensitivity reduction throughout the frequency spectrum for any viable control design \cite{skogestad2005multivariable}. Essentially, regardless of the control design, reducing the sensitivity at a lower frequency for asymptotic reference tracking, reducing the coupling effect, and synchronization will increase the sensitivity at other frequencies, causing a peak in the sensitivity transfer function, as shown in Fig. \ref{fig:Water_Bed}. The peak sensitivity $M_s = \|\widetilde{S}\|_\infty$, is directly proportional to the overshoot and oscillation in the transient response of the GFL inverter and is inversely correlated with the robustness of the closed-loop to the uncertainty of the plant. In the following proposition, we quantify how the performance of GFL places a theoretical limit on the minimum value that $M_s$ can achieve. 
\begin{proposition}
For any stable $\widetilde{S}$ in (\ref{eq:SISO_Sensitivity}), the peak sensitivity $M_s$ for each SISO loop $\widetilde{S}^q$ and $\widetilde{S}^q$ is lower-bounded by
{\small\begin{equation}
    \begin{split}
        \frac{1}{\omega_T - \omega_B}
        \left(\left|\int_{0}^{\omega_{B}}\ln{|\widetilde{S}^{\{d,q\}}|}d\omega\right|
        - \frac{3}{4}\omega_T \right)
        \leq \ln M_s,
    \end{split}
    \label{eq:peak_sensitivity_lower_bound}
\end{equation}}%
where $\omega_B$ and $\omega_T$ are shown in Fig. \ref{fig:Water_Bed} and defined as
{\small\begin{align}
    &\forall\; \omega \in \left[0,\omega_B\right],
    \;
    \ln{|\widetilde{S}|} \leq0,
    \label{eq:effective_control_bandwidth}
    \\
    &\forall\; \omega \in \left[\omega_T,\infty\right),
    \;
    \|\widetilde{G}_L K\left(j\omega\right)\|_2
    \leq \frac{1}{2}\left(
    \frac{\omega_{T}}{\omega}
    \right)^2.
    \label{eq:loop_upper_bound}
\end{align}}%
\end{proposition}
\begin{remark}
Equation (\ref{eq:peak_sensitivity_lower_bound}) connects the transient response and the stability margin in the form of the peak sensitivity $M_s$ to the low-frequency performance metrics, as shown in Fig. \ref{fig:Water_Bed}. Moreover, it presents all the knobs available to achieve a trade-off between conflicting control objectives. Note that
{\small\begin{equation}
    \begin{split}
        2\arcsin\left(\frac{1}{2M_s}\right)[\text{rad}] \leq \text{PM},\;
        \frac{M_s}{M_s-1} \leq \text{GM},
    \end{split}
    \label{eq:PM_and_GM}
\end{equation}}%
where to limit the oscillation and maintain the gain margin (GM) and phase margin (PM) of at least $6\text{dB}$ and $30^{\circ}$, it is sufficient to keep $M_s$ less than $2$\cite{skogestad2005multivariable}. Consequently, it is desirable to shape the sensitivity so that the left-hand side of (\ref{eq:peak_sensitivity_lower_bound}) becomes less than $\ln{2}$. Based on (\ref{eq:peak_sensitivity_lower_bound}) for any linear GFL control design, only four general and primary methods and their combination exist to reduce the peak sensitivity. These methods are\\
$\bullet$ {\bf Closed-loop bandwidth:} Increasing $\omega_T$, reduces the peak sensitivity by increasing both $0.75\, \omega_T$ and the denominator in (\ref{eq:peak_sensitivity_lower_bound}). Taking into account the definition of $\omega_T$ in (\ref{eq:loop_upper_bound}), the open-loop gain cross-over frequency is below $\omega_T$, and $\omega_T$ can be taken as a rough approximation of the closed-loop bandwidth. This approach leads to more noise and harmonics in the voltage of the capacitor and the inverter frequency.\\
$\bullet$ {\bf Effective control bandwidth:} Decreasing $\omega_B$ constrains the sensitivity integral limits in (\ref{eq:peak_sensitivity_lower_bound}) to a smaller frequency range while increasing the denominator. Based on (\ref{eq:effective_control_bandwidth}), $\omega_B$ represents the frequency range where $\widetilde{S}$ is less than one and the feedback reduces the effect of exogenous inputs (see (\ref{eq:Tracking_Error})); hence we call $\omega_B$ effective control bandwidth. We can reduce $\omega_B$ by decreasing the open-loop gain just below the gain cross-over frequency.\\
$\bullet$ {\bf Low-frequency performance:} The sensitivity integral is the main contributor to the peak sensitivity lower-bound in (\ref{eq:peak_sensitivity_lower_bound}). We can reduce the integral by increasing the sensitivity within the effective control bandwidth $\omega_B$. However, this sacrifices the quality of reference tracking, disturbance rejection, and synchronization. More explicitly, we can use following approximation of $\widetilde{S}$ within $[0,\omega_B]$  
{\small\begin{equation}
    \begin{split}
        \left|\int_0^{\omega_B}\ln{|\widetilde{S}^{\{d,q\}}|}d\omega\right|
        \approx
        \int_0^{\omega_B}\ln{|\widetilde{G}_L|} + 
        \ln{|K^{\{d,q\}}|} d\omega,
    \end{split}
    \label{eq:controller_integral}
\end{equation}}%
to capture the effect of the high-gain controller on the sensitivity integral. As is evident, PR harmonic compensators and high-gain pure integrators negatively impact the peak sensitivity by increasing above integral.\\
$\bullet$ {\bf Uniform sensitivity distribution:} Uniform distribution of $\widetilde{S}$ within the $[\omega_B,\omega_T]$ frequency range leads to smaller $M_s$ by making the inequality in (\ref{eq:peak_sensitivity_lower_bound}) tighter. We employed (\ref{eq:Upper_Bound_Second_Integral}) in the proof of the peak sensitivity inequality, and under the uniform distribution of $\widetilde{S}$ within $[\omega_B,\omega_T]$ this inequality changes to equality. Therefore, we can flatten the sensitivity transfer function by using the lead compensator within the $\left[\omega_B,\omega_T\right]$ frequency range and reduce $M_s$.
\end{remark}
\vspace{-4mm}
\section{Framework for Inverter Control Implementation and Synthesis}
\label{sec:Framework_for_Inverter_Control}
The controller $K$ in Fig. \ref{fig:q_control_split} incorporates the dynamics of the inverter from (\ref{eq:inverter_ss_dynamics}), which we have not taken into account in previous sections for a simplified analysis of performance and stability. In this section, we will first provide an example of a control design for $\{K^d,K_1^q,K_2^q\}$ that meets the stability and performance criteria discussed earlier. We will then demonstrate how the dynamics of the inverter and their associated closed-loop shape the controllers and their implementation.
\vspace{-5mm}
\subsection{Design of the Feedback Controllers}
We adopt the following control designs for $K^d$ and $K_1^q$:
{\small\begin{align}
    K^d &= 
    \left(
    \frac{L\omega_d \sqrt{\omega_d^2 + \lambda^2}}
    {\alpha_d s}
    \right)
    \left(
    \frac{s + \alpha_d \omega_d}{s + \omega_d/\alpha_d}
    \right),
    &\alpha_d &\leq 1,
    \label{eq:Kd_Design}
    \\
    K_1^q &=
    \left(
    \frac{L\omega_q^2}
    {\alpha_q}
    \frac{s + \alpha_q\omega_q}{s + \omega_q / \alpha_q}
    \right)
    \left(
    \frac{s}{(s + \lambda)(s + \omega_\theta / 5)}
    \right),
    &\alpha_q &\leq 1,
    \label{eq:K1q_Design}
    \\
    K_2^q &= 
    \left(
    \frac{L\omega_\theta^2}{\alpha_\theta s^2}
    \frac{s + \alpha_\theta\omega_\theta}{s + \omega_\theta/\alpha_\theta}
    \right)
    \left(
    \frac{s^2 + 2\lambda s + \lambda^2 + \omega_0^2}
    {s + \lambda}
    \right), 
    & \alpha_\theta &\leq 1, 
    \label{eq:K2q_controller}
\end{align}}%
where $L$ and $\lambda$ are same as (\ref{eq:GL_expression}).
$K^d$ in (\ref{eq:Kd_Design}) is composed of an integrator and a lead compensator and includes two tunable parameters $\{\alpha_d,\omega_d\}$. For $2\pi 120 \leq \omega_d$, the $d$ axis open-loop transfer function $K^d\widetilde{G}_L$ achieves a cross-over frequency close to $\omega_d$ and a phase margin close to
{\small\begin{align}
    90^{\circ} +
    \arcsin{\frac{1-\alpha_d^2}{1+\alpha_d^2}} -
    \arctan{\frac{\omega_d}{\lambda}}.
    \label{eq:d_axis_PM}
\end{align}}%
Subsequently, for the given $K^d$ in (\ref{eq:Kd_Design}), the sensitivity $\|\widetilde{S}^d\|$ as defined in (\ref{eq:SISO_Sensitivity}) is less than one below $\omega_d$, and also satisfies the decoupling condition in Proposition \ref{prop: decoupling}(c) and the stability condition in Proposition \ref{prop:stability}. Moreover, the integrator in $K^d$ automatically fulfills the zero steady-state tracking condition in Proposition \ref{prop:synchronization}(a). Note that smaller values of $\alpha_d$ result in robust stability and lower high frequency filter resonance.\\
$K_1^q$ in (\ref{eq:K1q_Design}) is composed of a lead compensator and a band-pass filter and includes three tunable parameters $\{\alpha_q,\omega_q,\omega_\theta\}$. $\omega_q$ and $\omega_\theta$ denote the desired cross-over frequency for $K^q\widetilde{G}_L$ and $K_2^q\widetilde{G}_L$ that specifies the bandpass frequency range $[\omega_\theta,\omega_q]$ for $\widetilde{T}_v^q$ as shown in Fig. \ref{fig:composit_response}. For $2\pi 120 \leq \omega_q$, the $q$ axis open-loop transfer function $K^q\widetilde{G}_L$ achieves a cross-over frequency close to $\omega_q$ and a phase margin close to
{\small\begin{align}
    \arcsin{\frac{1-\alpha_q^2}{1+\alpha_q^2}}.
    \label{eq:q_axis_PM}
\end{align}}%
The controller $K_2^q$ in (\ref{eq:K2q_controller}) includes two integrators and together with $K_1^q$ in (\ref{eq:K1q_Design}), they satisfies the zero steady-state tracking and synchronization objectives of Proposition \ref{prop:synchronization}. Furthermore, the proposed controller shapes the $K_2^q\widetilde{G}_L$ as
{\small\begin{align}
    K_2^q\widetilde{G}_L = 
    \frac{\omega_\theta^2}{\alpha_\theta s^2}
    \frac{(s + \alpha_\theta\omega_\theta)}
    {(s + \omega_\theta/\alpha_\theta)}.
    \label{eq: Inverter_Freq_Open_Loop}
\end{align}}%
Based on Proposition \ref{prop:freq_transient}, $K_2^q\widetilde{G}_L$ is the open-loop transfer function for $\widetilde{T}_\theta$. Therefore, based on Remark \ref{remark:Frequency_Transient}, the cross-over frequency $\omega_\theta$ is directly proportional to the closed-loop bandwidth of $\widetilde{T}_\theta$. Therefore, we achieve a more inertial response with smaller values of $\omega_\theta$. Additionally, increasing the phase margin by decreasing $\alpha_\theta$ reduces the frequency nadir (zenith) by damping the overshoot of the transient response.\\
To implement the proposed controllers, we have to understand the underlying embedding of the inverter dynamics and its closed-loop into the controllers. In what follows we use cascaded closed-loop to derive one such a relation.
\vspace{-3mm}
\subsection{Implementing Feedback Controller with Inverter Dynamics}\label{ControlSection}
\begin{figure}[t]
    \centering
    \subfloat[]{\includegraphics[width = 0.8\linewidth]{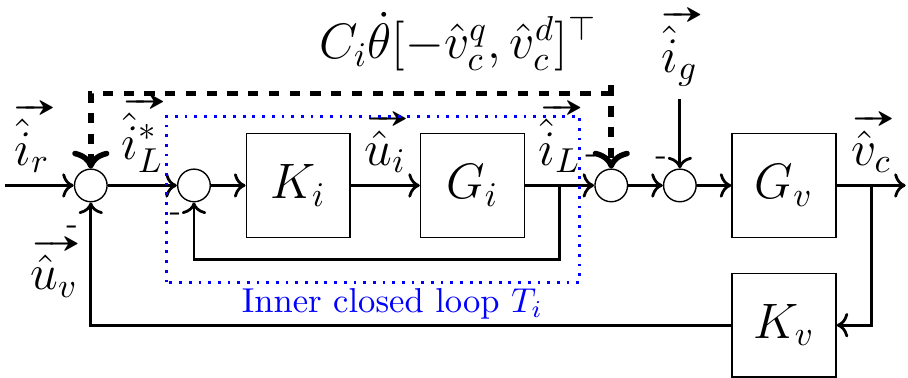}
    \label{fig:Inner_outer_voltage}}
    \\
    \vspace{-3mm}
    \subfloat[]{\includegraphics[width = 0.8\linewidth]{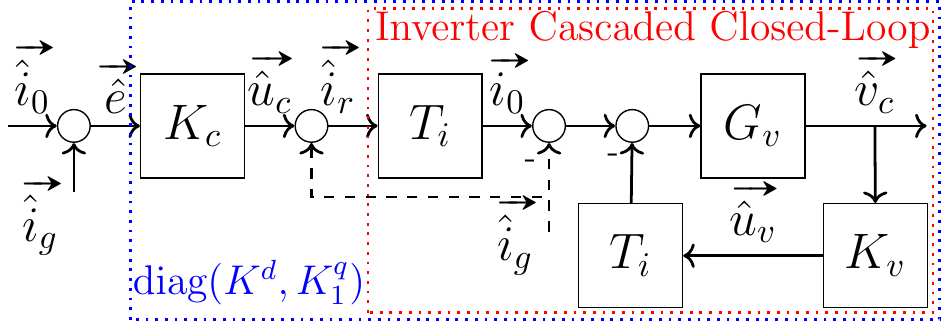}
    \label{fig:Control_Feedback_Form}}
    \caption{(a) Cascaded control structure with nested inner current closed-loop $T_i$ and outer voltage loop compensator $K_v$. (b) The algebraic equivalent of a nested architecture with series compensator $K_c$.}
    \vspace{-5mm}
\end{figure}

We adopt the cascaded architecture with inner current and outer voltage loops as shown in Fig. \ref{fig:Inner_outer_voltage} to form closed-loop around inverter dynamics in (\ref{eq:inverter_ss_dynamics}). Here $G_v$ and $G_i$ represent the inverter capacitor and inductor dynamics in (\ref{eq:inverter_ss_dynamics}). Additionally, to decouple the inductor dynamics in (\ref{eq:inverter_ss_dynamics}) we employ the feedback linearization by defining the modulation signal as
{\small\begin{equation}
    \begin{split}
        \vv{\hat{m}} = \frac{2}{v_{dc}} 
        \left(\vv{\hat{u}_i} + \vv{\hat{v}_c} + L_i \Dot{\theta}
        [
        -\hat{i}_L^q, \hat{i}_L^d
        ]^{\top} \right),
    \end{split}
    \label{eq:feedback_linearization}
\end{equation}}%
where $\vv{u}_i$ denotes the output of current compensator $K_i$ in Fig. \ref{fig:Inner_outer_voltage} and $v_{dc}$ is the inverter's input dc voltage. Subsequently,
we use $K_i = \left(L_i s + R_i\right)/\left(\tau_i s\right)$ to shape the inner closed-loop $T_i$ into a unity gain low-pass filter
{\small\begin{equation}
    \begin{split}
        T_i = 
        \frac{G_iK_i}{1 + G_iK_i}
        =
        \frac{1}{\tau_i s + 1},\quad
        S_i = 1 - T_i
        =
        \frac{\tau_i s}{1 + \tau_i s},
    \end{split}
    \label{eq:inner_closed_loop}
\end{equation}}%
where $1/\tau_i$ is the inner closed-loop bandwidth.\\
We close the outer voltage loop as shown in Fig. \ref{fig:Inner_outer_voltage} by setting the reference $\vv{i_L^*}$ for the inner closed-loop as 
{\small\begin{equation}
    \begin{split}
        \vv{\hat{i}_L^*} = \vv{\hat{i}_{r}} - \vv{\hat{u}_v}
        + C_i \Dot{\theta} [-v_c^q,v_c^d]^{\top}.
    \end{split}
\end{equation}}%
Here, $\vv{\hat{u}_v}$ is the output of the diagonal controller $K_v$, $C_i \Dot{\theta} [-v_c^q,v_c^d]^{\top}$ is the feed-forward term used to decouple the capacitor dynamics in (\ref{eq:inverter_ss_dynamics}), and $\vv{i_r}$ is the input to the nested closed-loop. Subsequently, the proposed nested inverter closed-loop results in the following relation between the inputs $\{\vv{i_r},\vv{i_g}\}$ and $\vv{v_c}$  
{\small\begin{align}
    \begin{split}
        \vv{\hat{v}_c} &= G_v S_v\left(T_i \vv{\hat{i}_{r}}
        - \vv{\hat{i}_g} - S_i C_i \Dot{\theta} [-v_c^q,v_c^d]^{\top}\right),\,\text{{\normalsize where}}
        \label{eq:inner_outer_loop}
    \end{split}\\
    \begin{split}
        S_v &=
        \begin{bmatrix}
        S_v^d & 0 \\ 0 & S_v^q
        \end{bmatrix}=
        \begin{bmatrix}
        (1 + G_v T_i K_v^d)^{\text{-}1} & 0\\
        0 & (1 + G_v T_i K_v^q)^{\text{-}1}
        \end{bmatrix}.
        \label{eq:voltage_sensitivity}
    \end{split}
\end{align}}%
Finally, we set $\vv{\hat{i}_r}$ in (\ref{eq:inner_outer_loop}) as $\vv{\hat{i}_r}= \vv{\hat{u}_c} + \vv{\hat{i}_g}$, where $\vv{u}_c$ denotes the output of diagonal compensator $K_c= \text{diag}(K_c^d,K_c^q)$. This transforms the nested architecture in Fig. \ref{fig:Inner_outer_voltage} into a natural feedback compensator shown in Fig. \ref{fig:Control_Feedback_Form} and given by
{\small\begin{align}
    \vv{\hat{v}_c} = G_vS_v(K_c T_i \vv{\hat{e}} 
    - S_i(\vv{\hat{i}_g} + C_i \Dot{\theta} [-v_c^q,v_c^d]^{\top}))
    \label{eq:exact_natural_feedback}
\end{align}}%
For small values of $\tau_i$, $S_i$ is negligible over a wide range of frequencies, and therefore, $S_i (\vv{\hat{i}_g} + C_i [-v_c^q,v_c^d]^{\top})$ is insignificant and we omit it from (\ref{eq:exact_natural_feedback}). This leads to the following feedback compensator
{\small\begin{align}
    \begin{bmatrix}
        \hat{v}_c^d \\
        \hat{v}_c^q 
    \end{bmatrix}
    =
    \begin{bmatrix}
        K^d & 0 \\
        0 & K_1^q 
    \end{bmatrix}
    \begin{bmatrix}
        \hat{e}^d \\ \hat{e}^q 
    \end{bmatrix},
    \,
    \begin{bmatrix}
        K^d & 0 \\
        0 & K_1^q 
    \end{bmatrix}
    =
    G_v T_i
    \begin{bmatrix}
        S_v^d K_c^d & 0
        \\
        0 & S_v^q K_c^q
    \end{bmatrix}.
    \label{eq:natural_feedback}
\end{align}}%
The equation above demonstrates the dependence of the controllers $K^d$ and $K_1^q$ on the inverter dynamics, which are represented by $G_v$, $T_i$, and $S_v$. $G_v$ and $T_i$ are already determined in (\ref{eq:inverter_ss_dynamics}) and (\ref{eq:inner_closed_loop}), while $S_v$ in (\ref{eq:voltage_sensitivity}) can only be shaped through the voltage compensator $K_v$. Consequently, the controllers $K^d$ and $K_1^q$ can only be adjusted by $K_c$ and $K_v$. Therefore, two main questions arise: which transfer functions $K^d$ and $K_1^q$ can represent, and how to find the corresponding $K_v$ and $K_c$. We answer these two questions considering the controller of the $d$ axis $K^d$, the analysis for the $q$ axis is the same. We can expand $K^d$ in terms of $K_c$ and $K_v$ as follows
{\small\begin{align}
    K^d
    =
    G_v T_i S_v^d K_c^d
    =
    \frac{D^d(s)}{C_i s (\tau_i s + 1)D^d(s) + N^d(s)}K_c^d,
    \label{eq:Kd_Expand}
\end{align}}%
where $N^d(s)$ and $D^d(s)$ are the numerator and denominator polynomials of voltage compensator $K_v = N^d(s)/D^d(s)$. Considering that $K_v^d$ and $K_c^d$ can represent any proper transfer function, the controller $K^d$ in (\ref{eq:Kd_Expand}) can only represent transfer functions of relative degree of at least two. We can relax this condition to include transfer functions with a relative degree of one if $\tau_i$ is small or equivalently if the inner current loop $T_i$ has a high enough bandwidth. In this case we can use the following approximation for (\ref{eq:Kd_Expand})
{\small\begin{align}
    K^d
    \approx
    \frac{D^d(s)}{C_i s D^d(s) + N^d(s)}K_c^d.
\end{align}}%
On the basis of above we can easily find $K_c$ and $K_v$ for any controller $\{K^d,K_1^q\}$ with relative degree of at least one. For example, the sample controllers $\{K^d,K_1^q\}$ in (\ref{eq:Kd_Design}) and (\ref{eq:K1q_Design}) are implemented using the following $K_c$ and $K_v$ controllers
{\small\begin{align}
    K_c^d &=
    \frac{\omega_d\sqrt{\omega_d^2 + \lambda^2}}
    {\alpha_d \omega_{LC}^2},
    &
    K_v^d &= 
    \frac{C_i\omega_d(1/\alpha_d - \alpha_d)s}
    {s + \omega_d \alpha_d},
    \\
    K_c^q &=
    \frac{\omega_q^2}
    {\alpha_q\omega_{LC}^2}
    \frac{s + \alpha_q\omega_q}{s + \omega_q / \alpha_q},
    &
    K_v^q &= C_i\left(\lambda + \frac{\omega_\theta}{5}\right) +
    C_i\frac{\lambda \omega_\theta}
    {5 s},
    \nonumber
\end{align}}%
where $\omega_{LC_i}=1/\sqrt{LC_i}$. As a final note, the proposed control design in this section subsumes the virtual impedance methods.%
\subsubsection*{Neglected Time Delay}
Till this point, we have neglected the time delay present in the inverter model. Generally, inverter dynamics is affected by two primary sources of delay: PWM delay, denoted as $T_{\text{PWM}}$, and delay due to DSP calculations, represented as $T_{\text{calc}}$ \cite{wang2021stability}. In this paper, we assume that the sampling frequency $f_{s}=1/T_s$ and the switching frequency $f_{sw} = 1/f_{sw}$ are the same. In this case, $T_{\text{PWM}} = 0.5T_s$, $T_{\text{calc}} = T_s$ and the total delay is $T_d = T_{\text{PWM}} +  T_{\text{calc}} = 1.5 T_s=1.5/f_s$. Time delay introduces an additional phase lag into the inverter's open-loop, which grows linearly with frequency. This lag can be quantified as $\phi_d = -T_d \omega = -1.5 (\omega
/f_s) \text{ [rad]}$. Note that by maintaining an appropriate phase margin for the $d$ and $q$ axis open-loops, as outlined in (\ref{eq:d_axis_PM}) and (\ref{eq:q_axis_PM}), we can effectively counteract the extra phase lag caused by time delay around the cross-over frequency. This approach improves the robustness of closed-loop stability against time delays.
\vspace{-3mm}
\section{Simulation and Experimental Results}
\begin{figure}
    \centering
    \includegraphics[width = 0.8\columnwidth]{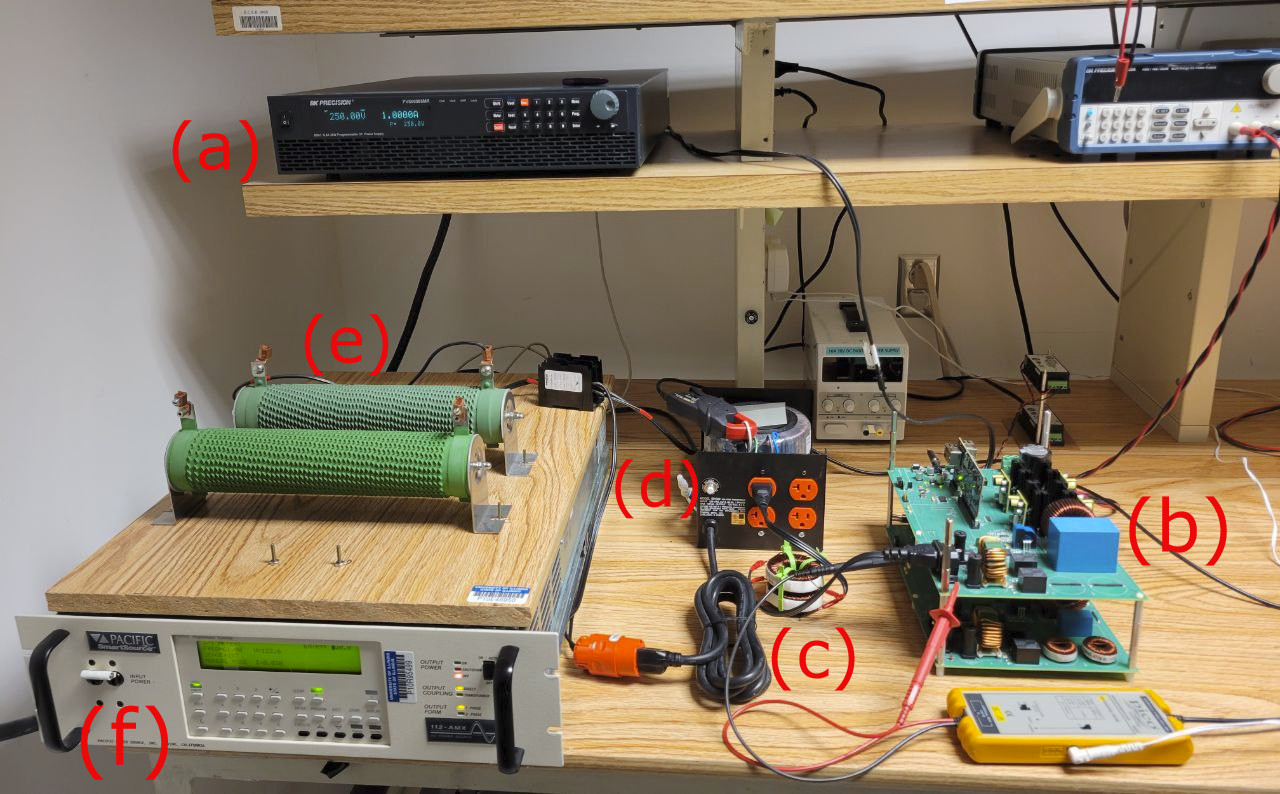}
    \caption{Experimental setup, (a) dc source (dc link), (b) inverters, (c) line inductors,(d) isolation transformer,(e) load,(f) AC source (grid simulator).}
    \label{fig:exp_setup}
    \vspace{-2mm}
\end{figure}

In this section, we demonstrate some of the salient features of the proposed framework through simulations and experiments. The simulations are carried out using the MATLAB Simscape toolbox; as for the experimental results, we used an inverter powered by a TI C2000 series DSP connected to a four-quadrant programmable AC source as a grid (Fig. \ref{fig:exp_setup}).
\begin{table}[ht]
    \vspace{-3mm}
    \centering
    \caption{Simulation and Experimental Parameters}
    \begin{tabularx}{\columnwidth}{>{\centering\arraybackslash}m{0.15\columnwidth}
                                     | >{\centering\arraybackslash}m{0.05\columnwidth}
                                     | >{\centering\arraybackslash}m{0.07\columnwidth} 
                                     | >{\centering\arraybackslash}m{0.09\columnwidth}
                                     | >{\centering\arraybackslash}m{0.18\columnwidth}
                                     | >{\centering\arraybackslash}m{0.05\columnwidth} 
                                     | >{\centering\arraybackslash}m{0.08\columnwidth}}
    \toprule[1pt]
    \midrule[0.3pt]
        Parameter & Sym. & Sim. & Expt. & Parameter & Sym. & Value \\
    \midrule
        Filter capacitor & $C_i$ & 50$\mu$F & 20$\mu$F & Grid voltage (line-RMS) & $v_g$ & 120V \\
    \hline
        Filter inductor & $L_i$ & 1mH & 3mH & Fundamental Frequency & $\omega_g$ & 60Hz\\
    \hline
        Grid inductor & $L$ & 1mH & 4.1mH & Switching frequency & $f_{sw}$ & 20kHz \\
    \hline
        Grid resistance & $R$ & 1m$\Omega$ & 1.4m$\Omega$ & Sampling Frequency & $f_s$ & 20kHz \\
    \midrule[0.3pt]
    \bottomrule[1pt]
    \end{tabularx}
    \label{tab:Sim_Expt_Parameter}
    \vspace{-3mm}
\end{table}\
\begin{figure}
    \centering
    \includegraphics[width = \columnwidth]{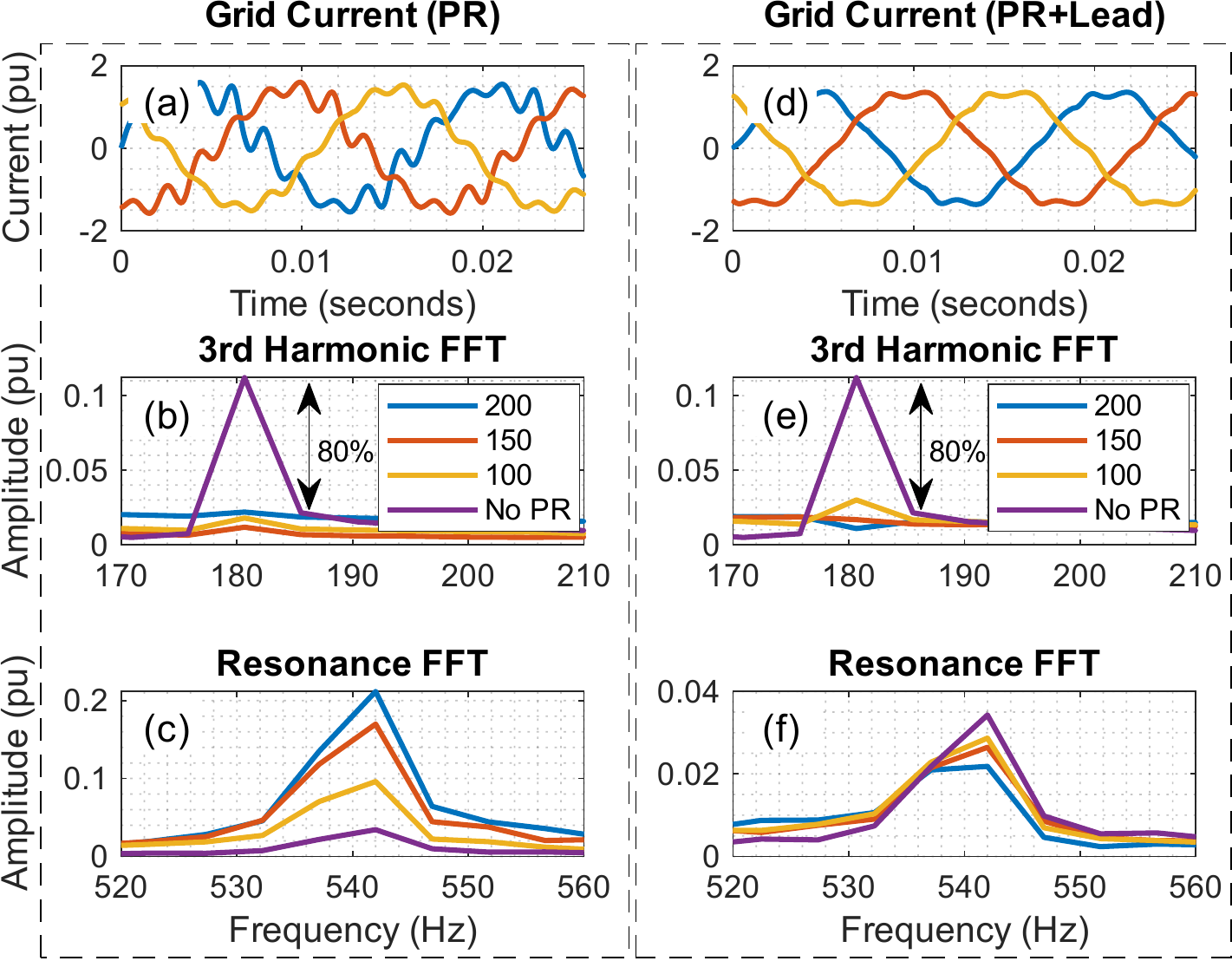}
    \caption{(a) PR induced current distortion. (b) PR compensators with distinct gains $\{100,150,200\}$ reduce harmonic content by 80\%, while (c) triggering filter resonance proportional to PR integral in (\ref{eq:controller_integral}). (d) PR + Lead scheme alleviates current distortion by (e) reducing the harmonic content by 80\% and (f) reducing PR-induced filter resonance.}
    \label{fig:PR_Resonance}
    \vspace{-5mm}
\end{figure}
\vspace{-5mm}
\subsection{Design for Harmonic Compensation vs. Filter Resonance}
Reducing $\|\widetilde{S}\|_\infty$ attenuates the effect of grid disturbance $\vv{d}$ on the output current $\vv{i_g}$, see (\ref{eq: GFL_Objectives_Laplace}) and (\ref{eq:disturbance_attenuation_bottleneck}). Our framework suggests a fundamental trade-off between reducing the sensitivity and increasing the peak resonance $M_s=\|\widetilde{S}\|_\infty$, see (\ref{eq:peak_sensitivity_lower_bound}). In practice, this shows the inherent limitation of popular PR schemes, where reducing $\widetilde{S}$ and attenuating the disturbance harmonics using the PR controller leads to an increase in peak sensitivity. An increase in $M_s$ induces filter resonance; This issue is typically addressed by tuning the PR parameters through trial and error. However, our framework provides theoretical ground, based on (\ref{eq:peak_sensitivity_lower_bound}), to improve the trade-off between harmonic attenuation and filter resonance.\\
To achieve a better trade-off between harmonic compensation and filter resonance, i.e. smaller $M_s$, we must improve the trade-off bound in (\ref{eq:peak_sensitivity_lower_bound}) by augmenting PR with an additional compensator. Based on (\ref{eq:peak_sensitivity_lower_bound}), reducing effective control bandwidth $\omega_B$, less aggressive low-frequency performance, and uniform sensitivity distribution help reduce peak sensitivity. These three objectives are achieved simultaneously by implementing a simple lead compensator of the form $(s + \alpha \omega_r)/(s + \omega_r/\alpha),\; \alpha < 1$, where $\omega_r$ is close to the resonant frequency.\\
To validate our approach, we used an AC source with 10\% third harmonic distortion. The standalone PR compensation is capable of reducing the third harmonic distortion by 80\%, as shown in Fig. \ref{fig:PR_Resonance}b. However, the standalone PR scheme induces the filter resonance in proportion to the PR gain illustrated in Fig. \ref{fig:PR_Resonance}c. The PR+Lead compensation scheme achieves the same third-harmonic attenuation as standalone PR, as shown in Fig. \ref{fig:PR_Resonance}e. However, the PR+Lead scheme reduces the filter resonance by at least 60\% in Fig. \ref{fig:PR_Resonance}f.
\begin{figure}[t]
    \centering
    \includegraphics[width = \columnwidth]{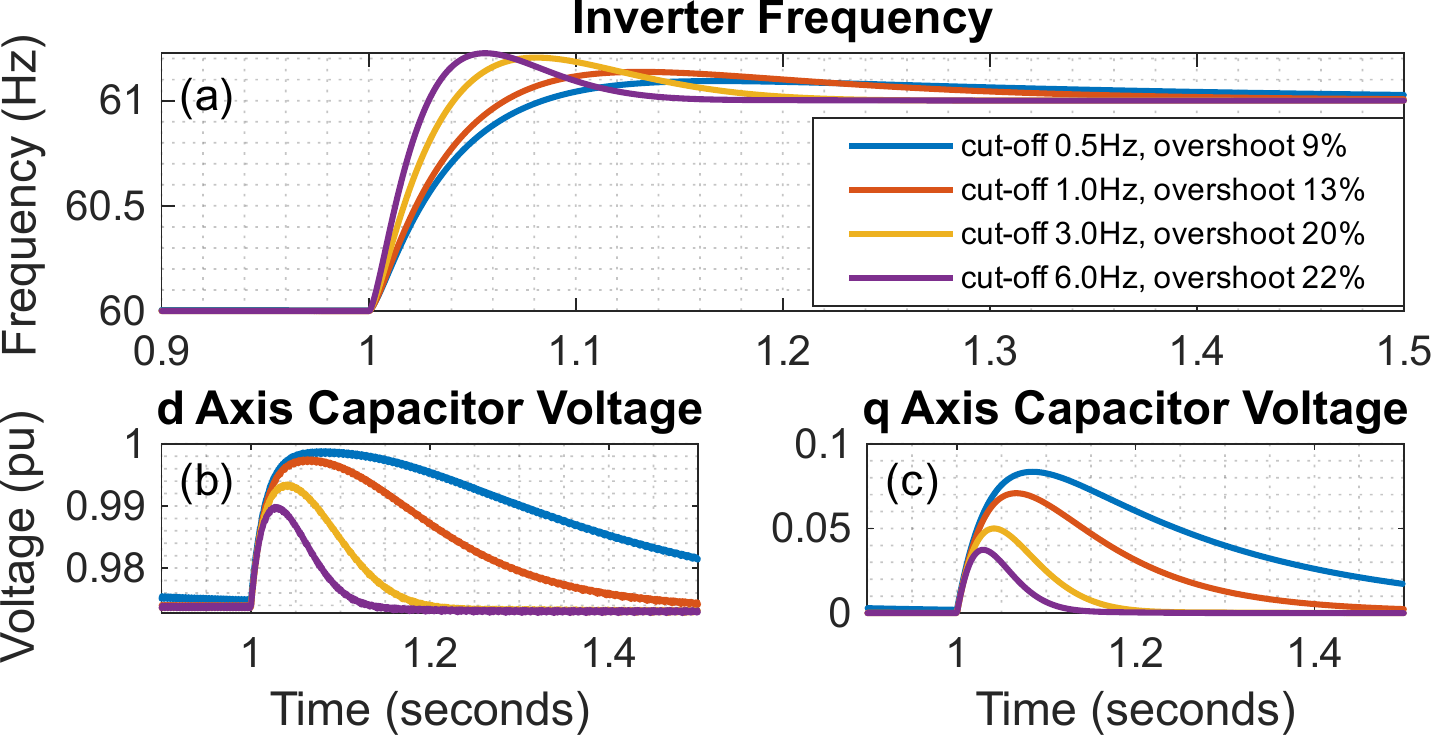}
    \caption{Transient response to $1$ Hz step change in grid frequency.}
    \label{fig:Freq_Dev}
    \vspace{-3mm}
\end{figure}
\begin{figure}[t]
    \centering
    \includegraphics[width = \columnwidth]{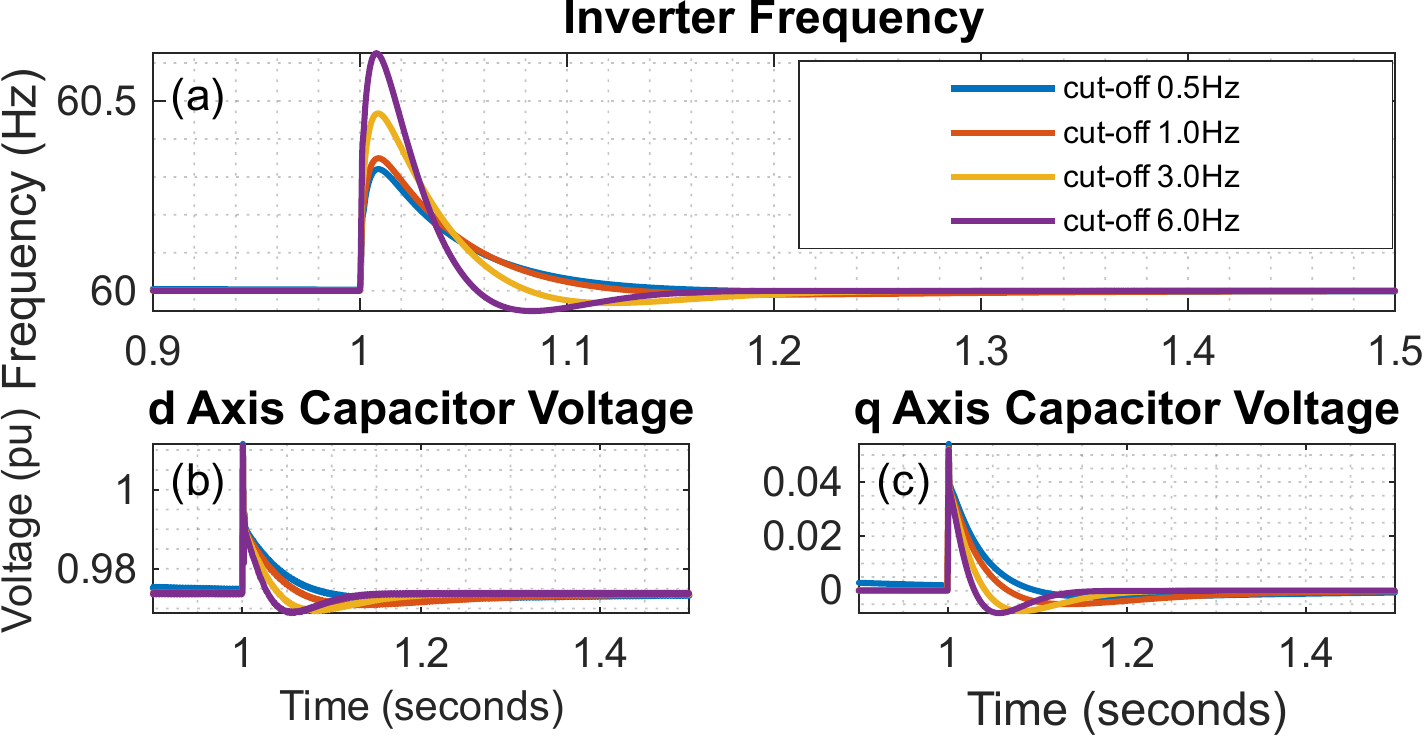}
    \caption{Transient response to $5^\circ$ step change in grid phase.}
    \label{fig:Phase_Dev}
    \vspace{-5mm}
\end{figure}
\begin{figure}[t]
    \centering
    \includegraphics[width = \columnwidth]{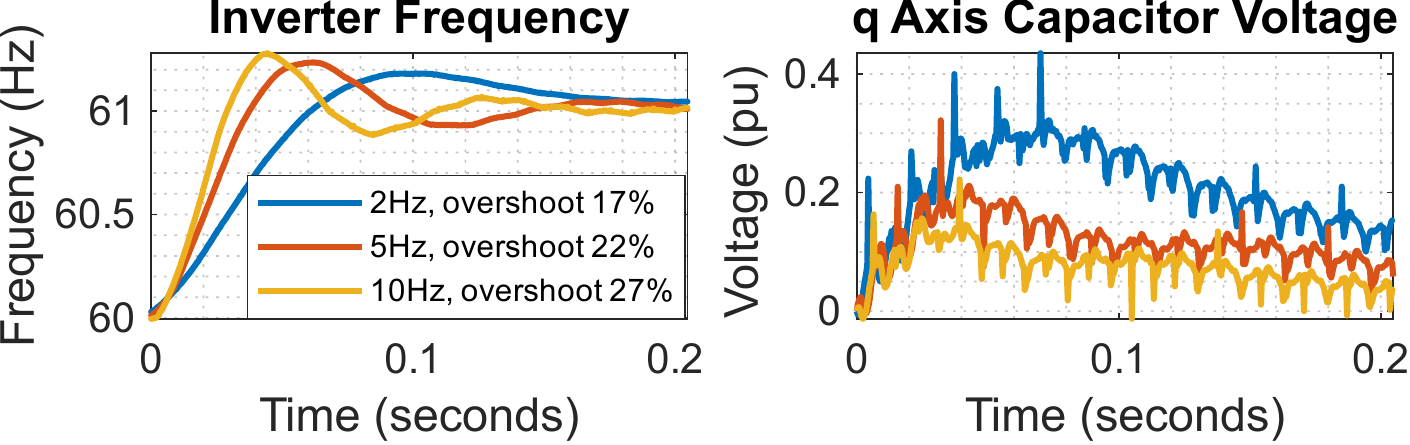}
    \caption{Transient response of experimental setup to $1$ Hz step change in grid simulator frequency.}
    \label{fig:Freq_Dev_Experiment}
    \vspace{-3mm}
\end{figure}
\begin{figure}
    \centering
    \subfloat[]
    {\includegraphics[width = \columnwidth]{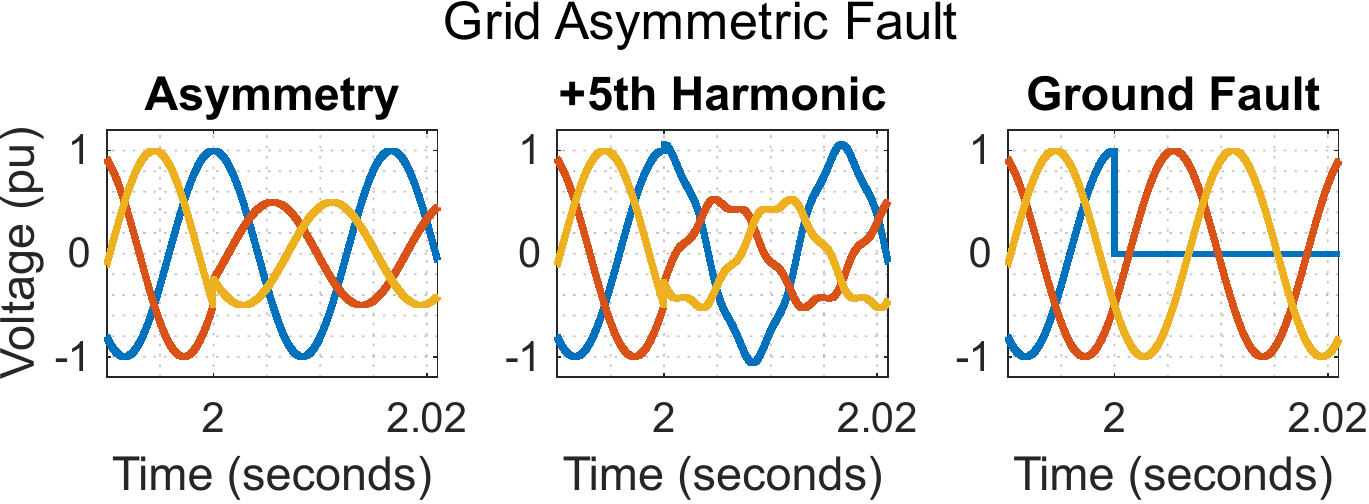}
    \label{fig:asymmetric_and_RS_a}}
    \\
    \vspace{-3mm}
    \subfloat[]
    {\includegraphics[width = 0.48\columnwidth]{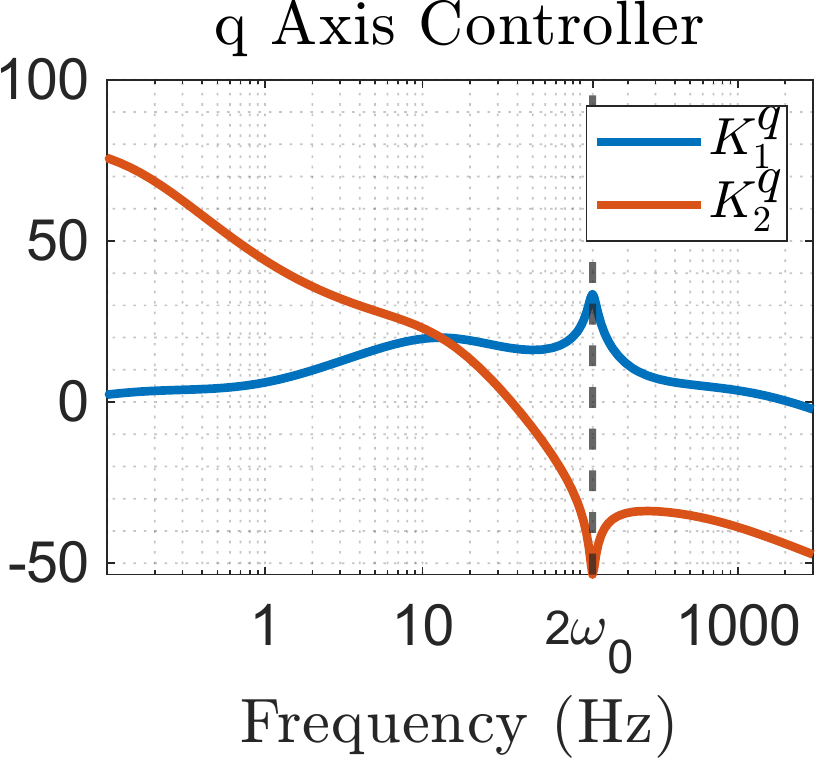}
    \label{fig:asymmetric_and_RS_b}}
    \subfloat[]
    {\includegraphics[width = 0.5\columnwidth]{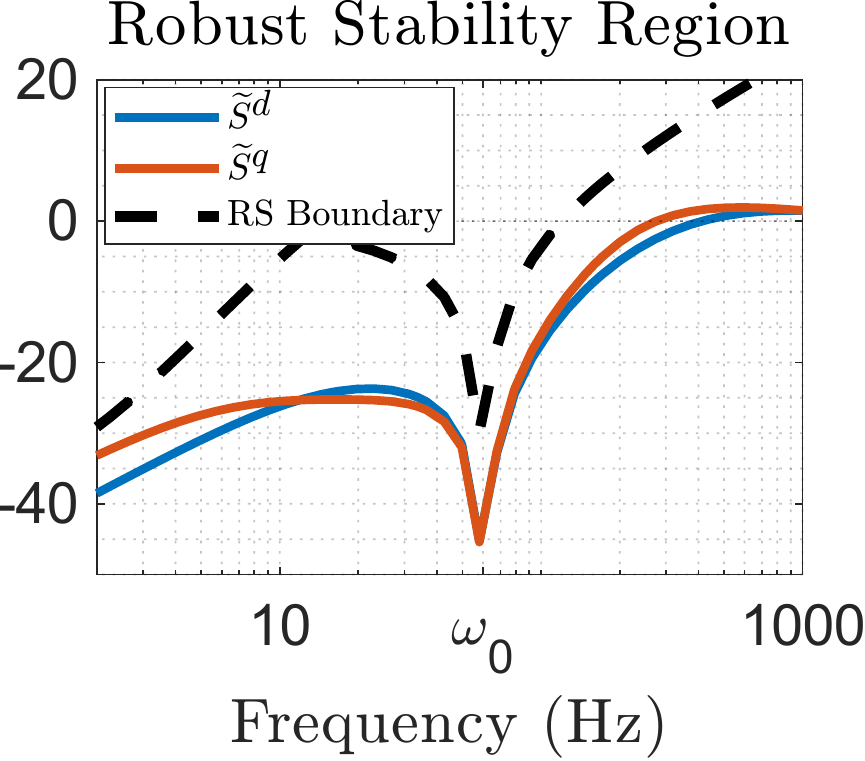}
    \label{fig:asymmetric_and_RS_c}}
    \caption{(a) The three asymmetrical faults investigated in this paper; Voltage imbalance, voltage imbalance with fifth harmonic distortion, and line-to-ground fault. (b) Parallel q-axis controller $K^q = K_1^q + K_2^q$ with $120$ Hz PR compensator in $K_1^q$ and $120$ Hz notch filter in $K_2^q$. (c) $d$ and $q$ closed-loop sensitivities of the E-GFL system, along with the robust stability boundary corresponding to the uncertainty sets in (\ref{eq: RS_Experiment_Values}).}
\end{figure}
\begin{figure}
    \centering
    \subfloat[]
    {\includegraphics[width = \columnwidth]{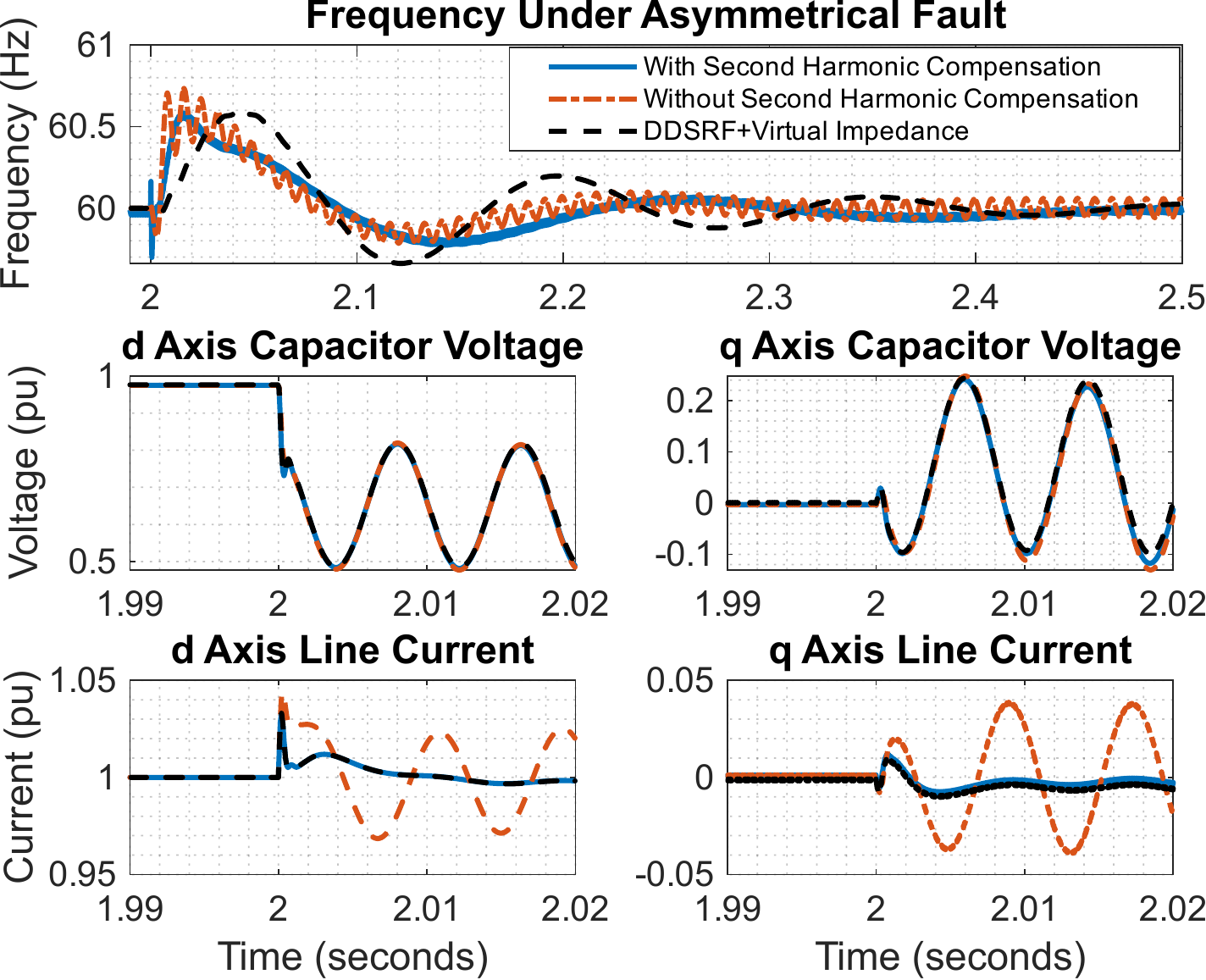}
    \label{fig:Ground_Fault_a}}
    \\
    \vspace{-3mm}
    \subfloat[]
    {\includegraphics[width = \columnwidth]{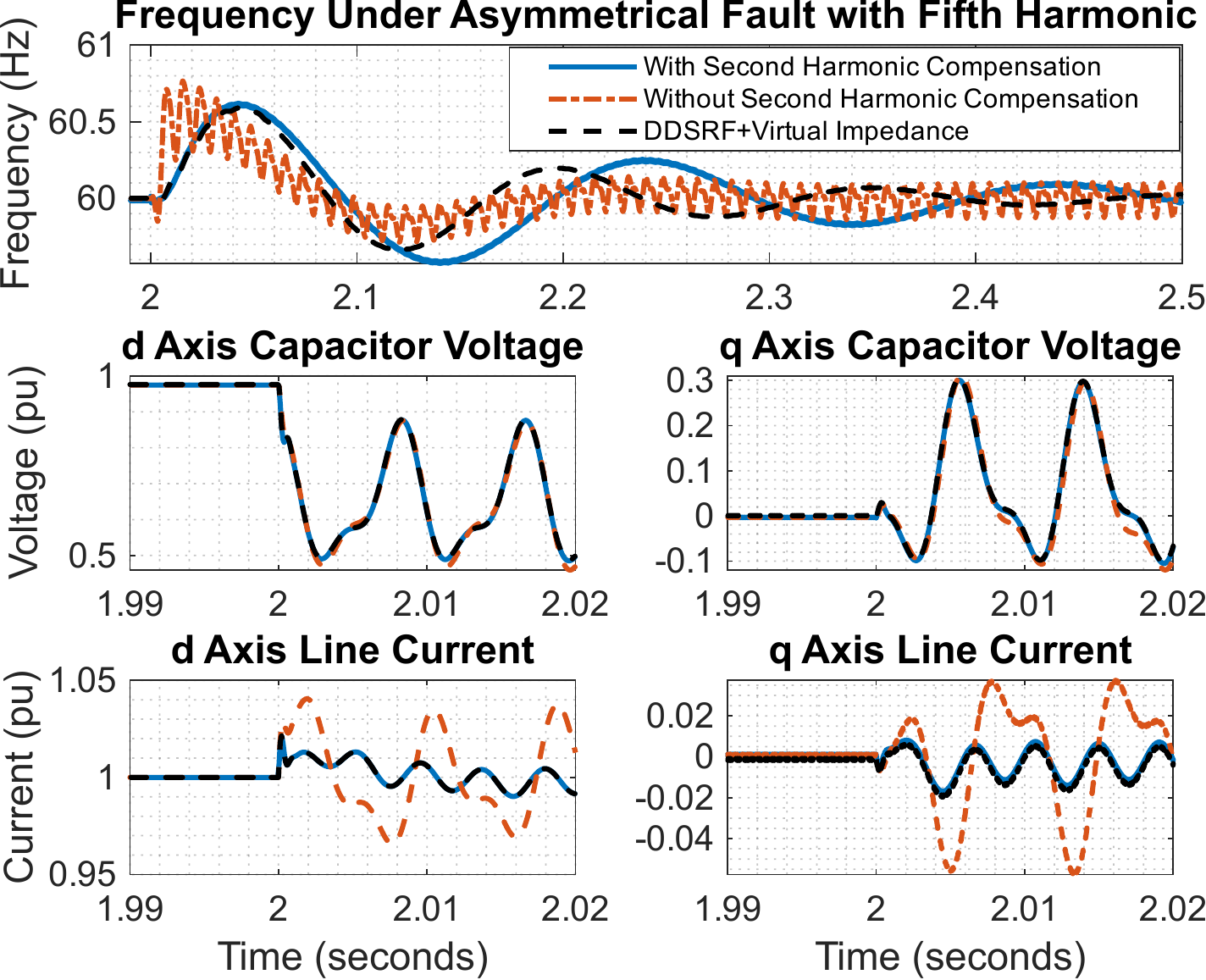}
    \label{fig:Ground_Fault_b}}
    \\
    \vspace{-3mm}
    \subfloat[]
    {\includegraphics[width = \columnwidth]{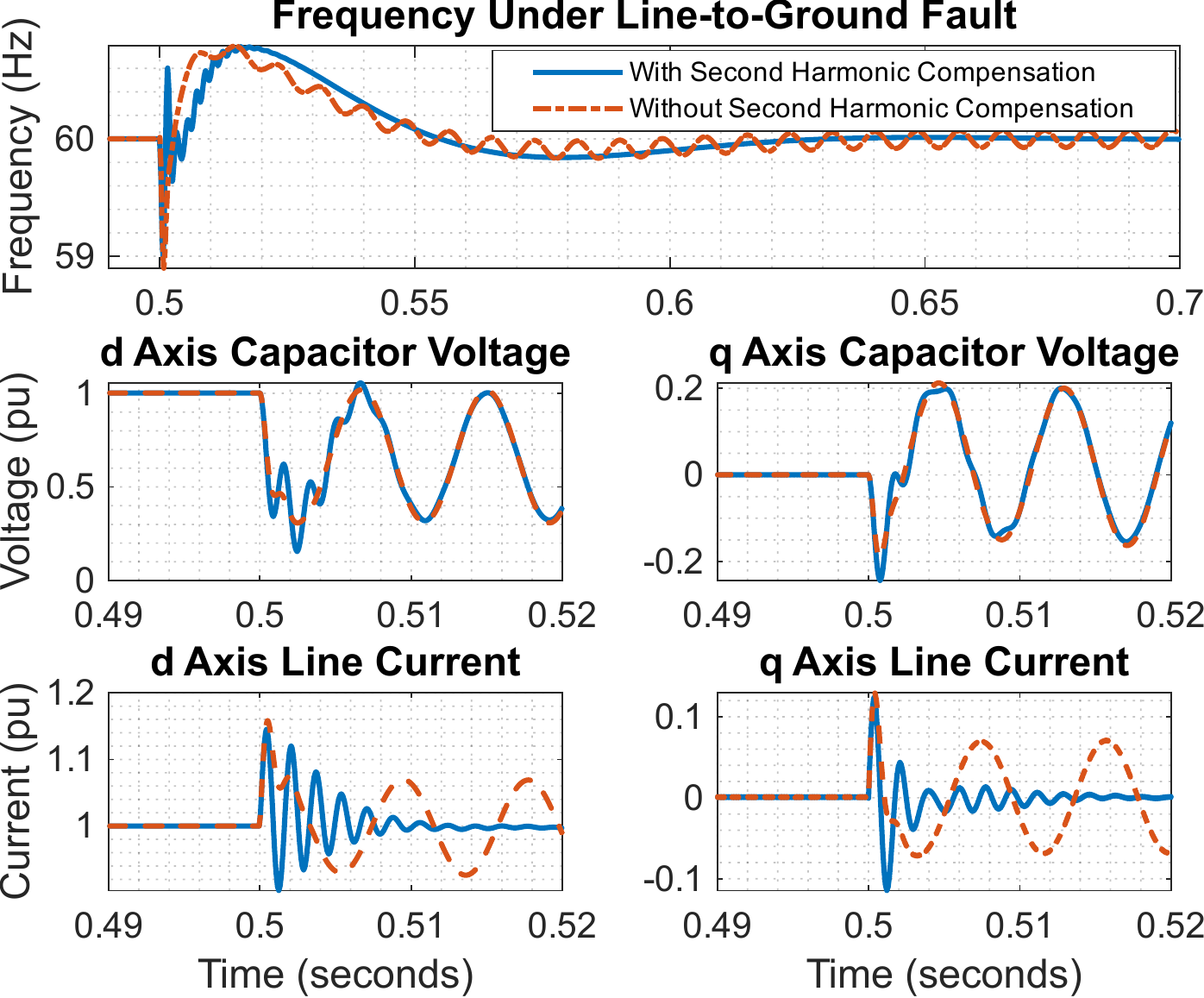}
    \label{fig:Ground_Fault_c}}
    \caption{(a) Asymmetrical fault, (b) asymmetrical fault with harmonic distortion, and (c) line-to-ground fault. The uncompensated closed-loop response (orange line) is provided for comparison.}
    \label{fig:Ground_Fault}
    \vspace{-5mm}
\end{figure}
\vspace{-3mm}
\subsection{Designing the Frequency Response To Grid Anomaly}
Weak grids are prone to phase and frequency jumps and fluctuations. This can negatively affect the inverter phase angle $\theta$ and distort the calculation of the $dq$ signals. In our framework, $d^q$ represents the grid frequency disturbance, and the algebraic constraint in (\ref{eq:Frequency_T_Bottleneck}) indicates the fundamental limitation where the effect of $d^q$ on $u_\theta$ and $v_c^q$ cannot be made small simultaneously. Therefore, as stated in Section \ref{remark:freq_volt_bottleneck}, it is desirable that $\widetilde{T}_{\theta}^q$ take the form of a unity gain low-pass filter that smooths the effect of grid frequency fluctuations on the inverter's frequency while $\widetilde{T}_v^q$ absorbs high frequency noise and harmonics (Fig. \ref{fig:composit_response}).\\
A series of simulations with different cut-off frequencies for $\widetilde{T}_{\theta}^q$ (dotted line in Fig. \ref{fig:composit_response}) demonstrate that a lower RoCoF and a slower transient response to a unit step change in the grid frequency (Fig. \ref{fig:Freq_Dev}a) and grid phase jump (Fig. \ref{fig:Phase_Dev}a) are obtained with lower cut-off frequencies. Furthermore, there is a noticeable decrease in the frequency overshoot (zenith) with lower cut-off frequencies. However, unlike the inverter phase and frequency, for lower cut-off frequencies the inverter voltage exhibits a larger excursion for the step change in the grid frequency (Figs. \ref{fig:Freq_Dev}b and \ref{fig:Freq_Dev}c) and grid phase jump (Figs. \ref{fig:Phase_Dev}b and \ref{fig:Phase_Dev}c). This is due to the constraint in (\ref{eq:Frequency_T_Bottleneck}). Finally, we performed the frequency step change in the experimental setup and observed that the inverter frequency and the $q$-axis voltage response in Fig. \ref{fig:Freq_Dev_Experiment} match the simulation results in Fig. \ref{fig:Freq_Dev}. Furthermore, the experimental results in Fig. \ref{fig:Freq_Dev_Experiment} show that most of the high-frequency distortion is shifted from $u_\theta$ to $v_c^q$. Note that the cutoff frequency of $\widetilde{T}_\theta^q$  in the classical sense is inversely correlated with the inertial type of response of the inverter, and we can control the frequency nadir by opting for lower cutoff frequencies.
\vspace{-3mm}%
\subsection{Design for Asymmetrical Fault}
Operating GFL inverters on an unbalanced grid presents significant challenges. The main issue is that the conventional PLL typically struggles to accurately lock onto the grid frequency and follow the positive sequence phase angle. The DDSRF PLL, as discussed in \cite{rodriguez2007decoupled}, effectively maintains phase lock with positive sequence even during asymmetric grid faults and can function amidst voltage imbalances distorted by high-order harmonics. However, this level of performance requires the use of two synchronous frames and nonlinear dynamics to properly separate the positive and negative sequences. A key advantage of our proposed control framework is its ability to match the performance of the DDSRF PLL during asymmetrical grid faults, but without the need for an additional synchronous frame or nonlinear dynamics. We accomplish this solely through the use of the PR compensator in the q-axis controller.\\
As extensively explained in Section \ref{sec: asymmetrical_fault}, asymmetric grid faults manifest as second-harmonic disturbances in the $dq$ frame. By integrating the PR controller, as specified in (\ref{eq:PR_Grid_Fault}), into $K^d$ and $K_1^q$, we can significantly diminish the impact of this second harmonic on both the output line current $\vv{i_g}$ and the $dq$ rotating angle $\theta$. Furthermore, the effects of asymmetric distortion on $dq$ angle $\theta$ can be further mitigated by cascading a notch filter at $2\omega_0$ with $K_2^q$. In Fig. \ref{fig:asymmetric_and_RS_b}, we illustrate the Bode magnitude plot for the components of the $q$ axis controller.\\
In this section, we compare our proposed control method with the DDSRF PLL's performance in three asymmetric fault scenarios. This comparison includes the DDSRF PLL as detailed in \cite{rodriguez2007decoupled} incorporated with a second harmonic compensator and an active virtual impedance damping with capacitor voltage feedback similar to that given in \cite{chen2023unified}. These enhancements are essential because, on its own, the DDSRF PLL lacks the necessary closed-loop stability margins and loop gain for stable functioning and efficient disturbance rejection of second harmonic.\\
$\bullet$ \textbf{First Scenario:} As shown in Fig. \ref{fig:asymmetric_and_RS_a}, two phases operate at 0.5 pu, and the third phase operates at nominal voltage. In this scenario, both our proposed method and the enhanced DDSRF PLL (with virtual impedance and second harmonic compensator) successfully maintain a clean output line current and a distortion-free $dq$ rotating frequency as shown in Fig. \ref{fig:Ground_Fault_a}.\\
$\bullet$ \textbf{Second Scenario:} This involves the same voltage imbalance as the first scenario, but with an additional fifth-order harmonic distortion as depicted in Fig. \ref{fig:asymmetric_and_RS_b}. Again, both our method and the DDSRF PLL manage to maintain a clean rotating frequency and effectively reject the second harmonic in the output current. However, due to the absence of a fifth-order PR compensator, some harmonic distortion remains in the output current, see Fig. \ref{fig:Ground_Fault_b}.\\
$\bullet$ \textbf{Third Scenario (Line-to-Ground Fault):} In this situation, one of the phases is directly connected to the ground, as shown in Fig. \ref{fig:asymmetric_and_RS_c}. In Fig. \ref{fig:Ground_Fault}, we observe that our methods again successfully achieve a clean rotating frequency and a distortion-free line current.\\
These comparisons underscore the effectiveness of our control method in addressing different types of asymmetric fault. The results indicate that the performance of our E-GFL method matches that of the DDSRF PLL, even when the DDSRF is augmented with extra measures such as virtual impedance active damping for increased stability margin. Notably, our method attains this level of performance without the need for an additional synchronous frame or nonlinear dynamics to separate the positive and negative sequences.
\begin{figure}
    \centering
    \subfloat[]
    {\includegraphics[width = \columnwidth]{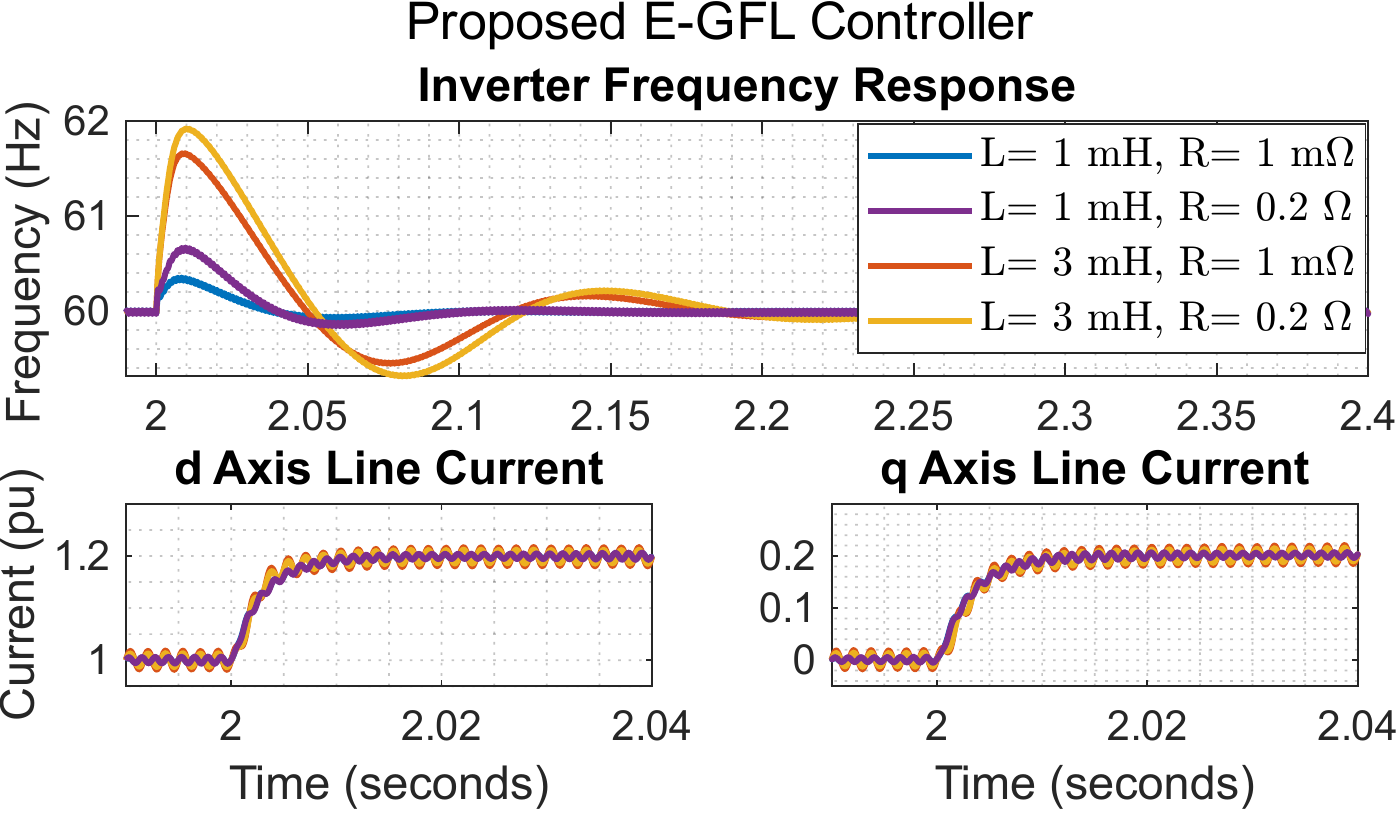}
    \label{fig: Line_Uncertainty_Experiment_a}}
    \\
    \vspace{-3mm}
    \subfloat[]
    {\includegraphics[width = \columnwidth]{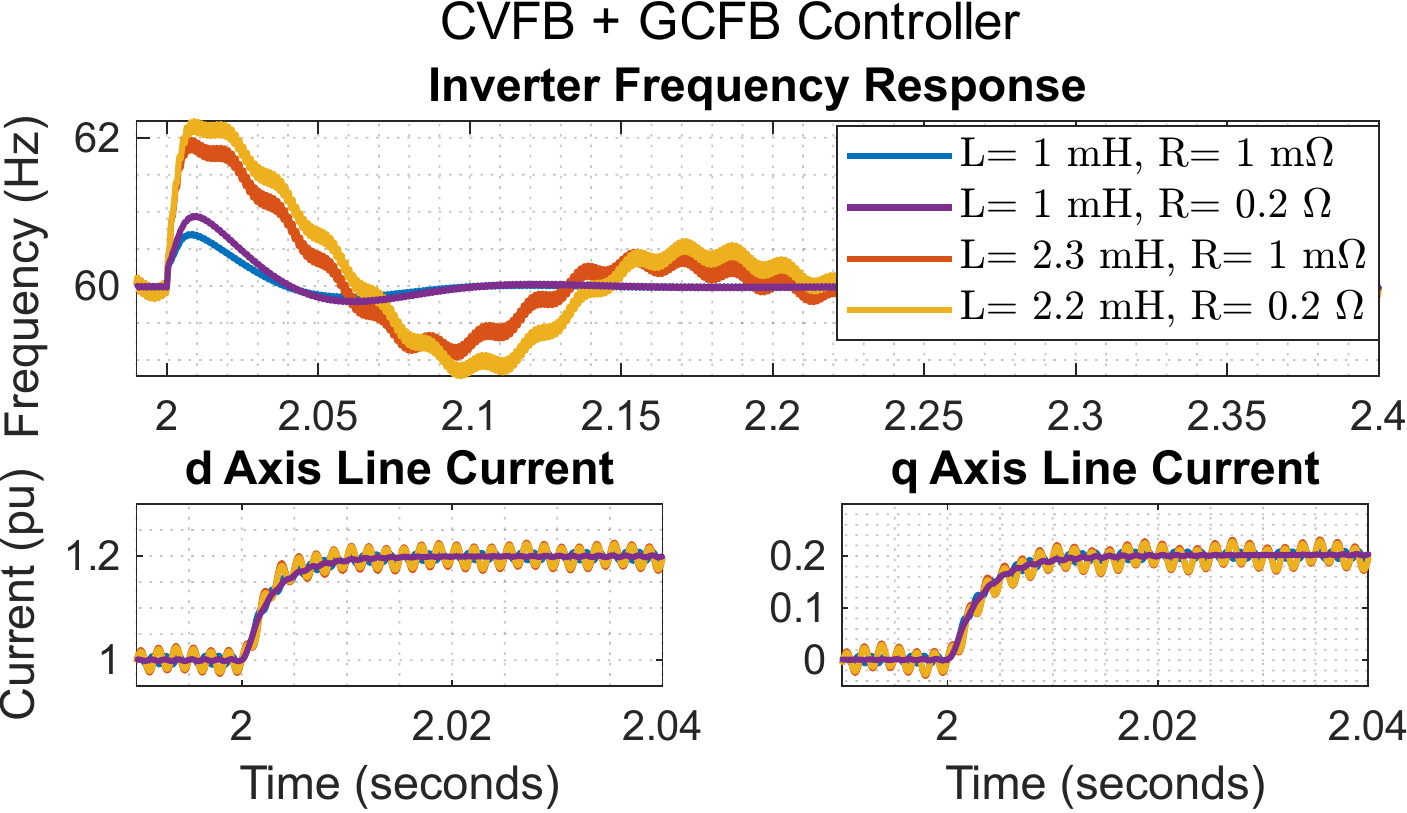}
    \label{fig: Line_Uncertainty_Experiment_b}}
    \caption{(a) The proposed E-GFL approach demonstrates its current tracking capabilities and rotating frame frequency response when faced with uncertainties in line impedance. (b) The current tracking performance and frequency reponse of CVFB+GCFB in face of line impedance uncertainty.}
\end{figure}
\vspace{-4mm}
\subsection{Robust Stability Under Line Parameter Uncertainty}
In Section \ref{sec: SISO_Robust_Stability}, we outlined the conditions required for robust stability when operating with uncertain, yet bounded, line impedance. To confirm these conditions, as proposed in Proposition \ref{prop: robust_stability}, we examine the closed-loop performance and stability for the following sets of uncertainties in line inductance and resistance
{\small\begin{align}
    L \in
    \Pi_L = [1 \text{ mH},3 \text{ mH}],
    \quad
    R \in
    \Pi_R = [1 \text{ m}\Omega , 200 \text{ m}\Omega].
    \label{eq: RS_Experiment_Values}
\end{align}}%
Considering the uncertainty sets above and referring to (\ref{eq:nominal_RS_parameters}), the nominal parameters for control design are as follows: inductance $L_0 = 1.5$ mH, $R_0 = 225 \text{ m}\Omega$, and $\lambda_0 = 150$. Using these nominal values along with the robust stability conditions outlined in (\ref{eq:first_RS_condition}) and (\ref{eq:second_RS_condition}), we established the Robust Stability (RS) boundary, depicted by the dashed line in Fig. \ref{fig:asymmetric_and_RS_c}. To ensure robust stability, the magnitudes of the SISO sensitivities $\widetilde{S}^d$ and $\widetilde{S}^q$ must remain below this RS boundary. We have successfully designed a controller such that the 2-SISO sensitivities, as shown in Fig. \ref{fig:asymmetric_and_RS_c}, stay within the confines of the RS boundary. Simulations demonstrate that these controllers maintain stability as the line impedance varies within the uncertainty set. In Fig. \ref{fig: Line_Uncertainty_Experiment_a}, we present the current reference tracking performance and the inverter frequency response when the step change of $0.2$ pu is applied to both $d$ and $q$ current set-points. This performance is observed at the four extreme points of line impedance variation, namely, the four possible combinations of the highest and lowest values of line inductance and resistance.\\
To assess the robustness of our proposed E-GFL controller, we compared it with a GFL inverter that uses standard PLL structure, as shown in Fig. \ref{fig:PLL_Conventional} and a combination of capacitor voltage feedback (CVFB) and grid current feedback (GCFB) \cite{chen2023unified} for active damping and stabilization. For fairness in comparison, we used the same nominal line values ${L_0, R_0, \lambda_0}$ to adjust the control design parameters in both setups. The results are shown in Fig. \ref{fig: Line_Uncertainty_Experiment_b}, reveal that the CVFB+GCFB approach performs comparably to the E-GFL at a lower inductance value of $L = 1$ mH. However, it is observed that the GFL inverter approaches the brink of instability when the inductance reaches approximately 2.3 mH. 
This finding implies that our proposed method provides a better margin of robustness. Specifically, the closed-loop can sustain stability across a range of line inductances approximately 1.5 times greater than that managed by the CVFB+GCFB method. This factor is calculated as $(3\text{ mH} - 1\text{ mH})/(2.3\text{ mH} - 1\text{ mH}) \approx 1.5$.
\begin{figure}
    \centering
    \subfloat[]
    {\includegraphics[width = \columnwidth]{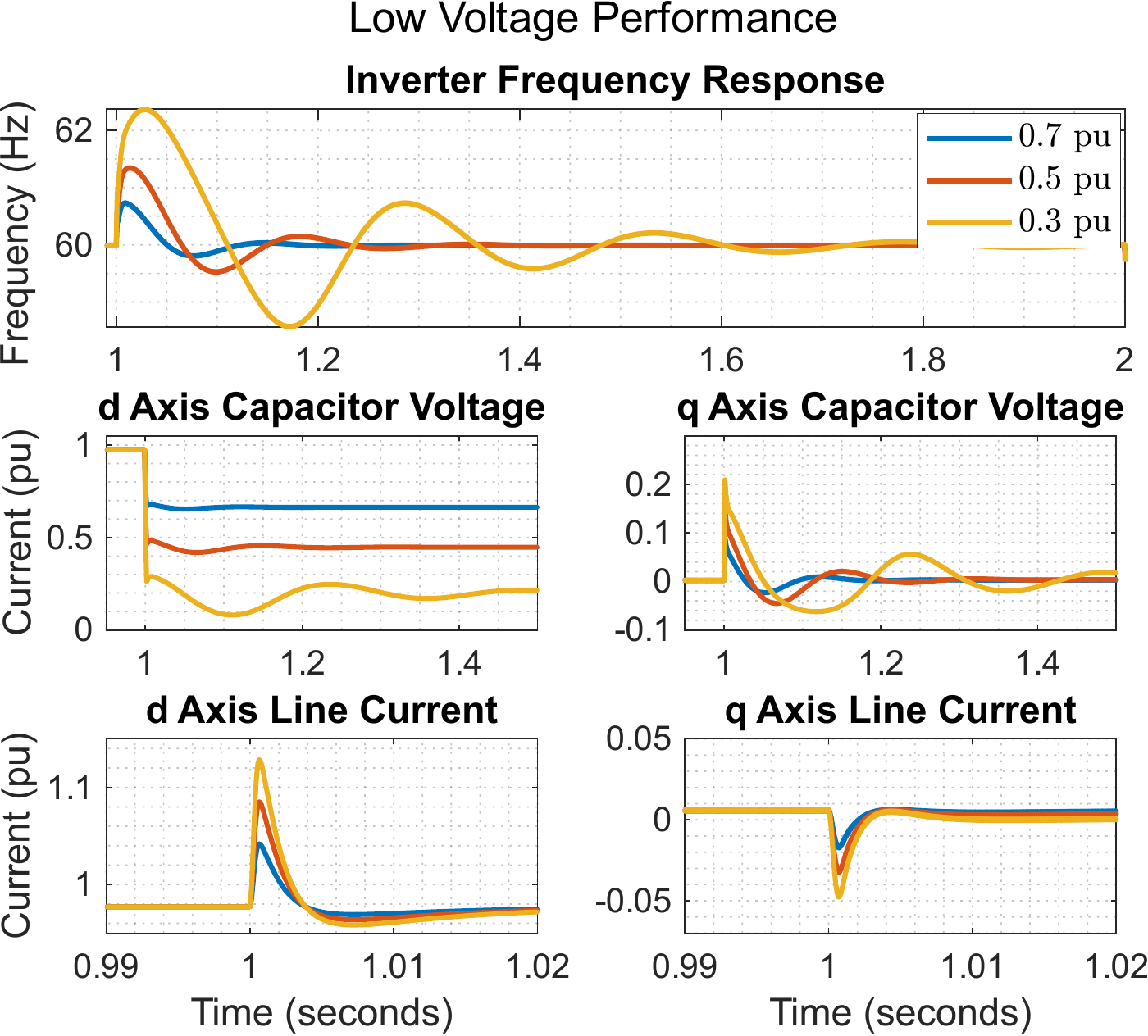}}
    \caption{Our control design successfully provides Low Voltage Ride Through (LVRT) capability down to 30\% of the nominal voltage value. However, as the voltage dips, the stability margins decrease, which is evident from the increasing oscillations in the system's transient response.}
    \label{fig:LVRT}
\end{figure}
\subsection{Low Voltage Performance}
During peak power demand, the grid may undergo substantial voltage dips. Such voltage sags can inadvertently cause the GFL inverters to disconnect, triggering a chain reaction in which the DERs are progressively cut off from the grid. This can worsen the situation by further destabilizing the power balance and deepening the voltage drop. Therefore, it is crucial for GFL inverters to have Low Voltage Ride-Through (LVRT) capability. Our proposed control design successfully achieves LVRT, maintaining operation even when the grid voltage falls to as low as 30\% of its nominal value. Fig. \ref{fig:LVRT} illustrates how the proposed controller performs at three different levels of voltage drop. As the voltage dip intensifies, the transient response of the closed-loop system becomes increasingly oscillatory. It is also important to note that the proposed controller starts to lose stability when the voltage drops close to 25\% or less of its nominal value.
\section{Conclusion}
We presented a novel and comprehensive framework for GFL control that exploits the algebraic structure of the MIMO line dynamics and captures the effect of coupled dynamics on the stability and performance of the closed-loop. The proposed framework embeds the inverter dynamics in the feedback controller and links different performance criteria, such as reference tracking, synchronization, harmonic attenuation, and filter resonance. In future work, we will extend the proposed framework to GFM inverters to accommodate a seamless transition between the GFL and GFM modes of operation.
\vspace{-5mm}
\section{Appendix}
\begin{proof}[Proposition 3.1]
The inverter voltage $\vv{v_c}$ is controllable and is part of $\vv{u}$. The grid voltage $\vv{v_g}$ is not controllable or even measurable. However, in the $dq$ frame we have 
{\small\begin{equation}
    \begin{split}
        \vv{v_g}= 
        \begin{bmatrix}
        v_g^d \\ v_g^q
        \end{bmatrix} =
        \begin{bmatrix}
        \|{v_g}\|\cos(\delta) 
        \\
        \|{v_g}\|\sin(\delta)
        \end{bmatrix} = \begin{bmatrix}
        \|{v_g}\|\hphantom{\delta} 
        \\
        \|{v_g}\|\delta
        \end{bmatrix}
         + \mathcal{O}(\delta^2),     
    \end{split}
    \label{eq:vg_expansion}
\end{equation}}%
where $\delta=\theta_g - \theta$ is the angle between the grid voltage phasor and the rotating frame as shown in Fig. \ref{fig:DQ_Frame}. $v_g^q$ depends on the orientation of the rotating frame through $\delta$, while $v_g^d$ is close to the magnitude of the grid voltage and is independent of the rotating frame. We exploit dependence on orientation and the small angle approximation $\mathcal{O}\left(\delta^2\right)\approx 0$ to rewrite $v_g^q$ as
{\small\begin{equation}
    \begin{split}
        v_g^q = 
        \|{v_{g}}\|\delta = 
        \underbrace{\|{v_g}\| \int (\Dot{\theta}_g - \omega_0)\;dt}
        _{\text{\large $d^{q}$}} 
        - \underbrace{\|{v_g}\| \int (\Dot{\theta} - \omega_0)\;dt}
        _{\text{\large $u_\theta$}},
    \end{split}
    \label{eq:vq_disturbance_control}
\end{equation}}%
splitting $v_g^q$ into disturbance $d^q$, and manipulated variable $u_\theta$. However, since we do not have the measurement of $v_g$, we approximate it by $v_0$ in calculating $u_\theta$.$\blacksquare$
\end{proof}%

\begin{proof}[Proposition 3.2]
(a) Note that
{\small\begin{equation}
    \begin{split}
        S =
        (I_{2} + \widetilde{G}_{L}K
        )^{\text{-}1}
        (I_{2} + E(I_{2} - \widetilde{S})
        )^{\text{-}1}.
        \label{eq:sensitivity_factorization}
    \end{split}
\end{equation}}%
By definition, we replace $E$ with {\small$\left(\Gamma^{\text{-1}} - I_2\right)$} in (\ref{eq:sensitivity_factorization}) and factor out $\Gamma^{\text{-}1}$ to get (\ref{eq:sensitivity_gamma}).$\blacksquare$ \\ 
(b) Replace the sensitivity factorization (\ref{eq:sensitivity_gamma}) into (\ref{eq:Tracking_Error}), and (\ref{eq:Control_Effort}).
(c) We use the Neumann series to get
{\small\begin{equation}
    \mathcal{X}_c = 
    (I_2 + (\Gamma - I_2)\widetilde{S}
    )^{\text{-}1}
     = I_2 +
    \sum_{n=1}^\infty (-(
    \Gamma - I_2
    )\widetilde{S})^n.
    \nonumber
\end{equation}}%
Hence $\|\mathcal{X}_c - I_2\|$ satisfies the following bound
{\small\begin{equation}
    \begin{split}
        \|\mathcal{X}_c - I_2\|_2 
        &\leq
        \sum_{n=1}^\infty \left(\|
        \Gamma\left(j\omega\right) - I_2
        \| 
        \|\widetilde{S}\left(j\omega\right)\|_2\right)^n
        = \frac{\epsilon\left(\omega\right)}{1 - \epsilon\left(\omega\right)}
        .\; \blacksquare
        \nonumber
    \end{split}
\end{equation}}%
\end{proof}%

\begin{proof}[Proposition 4.1]
Both $G_L$ and $K$ are stable transfer functions (we only consider stable controllers); hence the closed-loop in Fig. \ref{fig:q_control_split} is internally stable if and only if $S$ is stable. Stability of $\widetilde{S}$ and $\mathcal{X}_c$ in (\ref{eq: cross_mimo_dynamic}) Guarantee the stability of $S$ ($\Gamma$ is always stable). The first condition assumes stability of $\widetilde{S}$, and the second condition provides the following spectral radius upper bound
$
\rho( (\Gamma\left(j\omega\right)-I_{2} )
\widetilde{S} (j\omega) ) \leq
\epsilon (\omega) < 1,\; \forall\omega,
$
which is the sufficient condition for stability of $\mathcal{X}_c$. $\blacksquare$
\end{proof}%

\begin{proof}[Proposition 4.2]
(a) We have $\widetilde{G}_{L0} \in \Pi_G$. Therefore, if $\widetilde{S}_0$ in (\ref{eq:nominal_plant_sensitivity}) is unstable, then there is at least one plant in the set $\Pi_G$ that leads to instability and violates the RS condition.\\
(b) We can ensure robust stability if conditions (a) and (b) of Proposition \ref{prop:stability} are satisfied for all perturbed SISO plants $\widetilde{G}_{Lp} \in \Pi_G$. To fulfill condition (a), note that the sensitivity for an arbitrary $\widetilde{G}_{Lp} \in \Pi_G$ is
{\small\begin{align}
    \widetilde{S}_{p} =
    (I_2 + K \widetilde{G}_{Lp})^{\text{-}1}.
    \nonumber
\end{align}}%
We can always represent $\widetilde{S}_{p}$ for any $\widetilde{G}_{Lp}\in \Pi_G$ in terms of the nominal plant $\widetilde{G}_{L0}$ in (\ref{eq:nominal_RS_Plant}) as 
{\small\begin{align}
     \widetilde{S}_{p} =
    \left(
    I_2 + K \widetilde{G}_{L0}
    \frac{1 + \Delta_1 W_1}{1 + \Delta_2 W_2}
    \right)^{\text{-}1},
\end{align}}%
where $(1 + \Delta_1 W_1)$ represents the uncertainty in zeros of the plant and $(1 + \Delta_2 W_2)$ expresses the uncertainty in poles of the plant. The transfer functions $\Delta_1 (s)$ and $\Delta_2 (s)$ represent the norm-bounded complex perturbations that satisfy
{\small\begin{align}
    \|\Delta_1(j\omega)\|_\infty &\leq 1,
    &
    \|\Delta_2(j\omega)\|_\infty &\leq 1.
    \nonumber
\end{align}}%
and the uncertainty weights $W_1(s)$ and $W_2(s)$ for the set of perturbed plants $\Pi_G$, are any transfer function that satisfy
{\small\begin{align}
    \| W_1(j\omega)\|_2 & \geq \max_
    {\substack{
    \lambda \in \Pi_\lambda \\ L \in \Pi_L
    }}
    \left\{
    \left\|
    \frac{(j\omega + \lambda)/L - (j\omega + \lambda_0)/L_0}
    {(j\omega + \lambda_0)/L_0}
    \right\|_2
    \right\}, \; \forall \omega,
    \label{eq:w1_w2_weights}
    \\
    \| W_2 (j\omega) \|_2 & \geq \max_{\lambda \in \Pi_\lambda}
    \left\{
    \left\|
    \frac{(2\lambda j\omega + \lambda^2) - 
    (2\lambda_0 j\omega + \lambda_0^2)}
    {2\lambda_0 j\omega + \lambda_0^2 + \omega_0^2 - \omega^2}
    \right\|_2
    \right\}, \; \forall \omega.
    \nonumber
\end{align}}%
Based on robust control theory \cite{skogestad2005multivariable}, the perturbed SISO sensitivity $\widetilde{S}_{p}$ is stable for all $\widetilde{G}_{Lp} \in \Pi_G$ if
{\small\begin{align}
    \|\widetilde{S}_0 W_2\|_2 + 
    \|(I_2 - \widetilde{S}_0) W_1\|_2 < 1,
    \; \forall \omega,
    \label{eq:RS_Condition}
\end{align}}%
where $\widetilde{S}_{0}$ is the nominal sensitivity in (\ref{eq:nominal_RS_Plant}) and $W_1$ and $W_2$ are any transfer function that satisfy (\ref{eq:w1_w2_weights}). Although the choices for $W_1$ and $W_2$ in (\ref{eq:w1_w2_weights}) are not unique, our proposed weights in (\ref{eq:w1_design}) and (\ref{eq:w2_design}) lead to a less conservative robust stability bound in (\ref{eq:RS_Condition}). Finally, we use the following approximation to further simplify the RS condition in (\ref{eq:RS_Condition})  
{\small\begin{align}
    I_2 - \widetilde{S}_0 \approx 
    \frac{\omega_{bw}}{s + \omega_{bw}} = W_3
    \xrightarrow{}
    \|\widetilde{S}_0 W_2\|_2 < 1 - 
    \|W_1 W_3\|_2,\; \forall \omega.
    \nonumber
\end{align}}%
Therefore, satisfying (\ref{eq:first_RS_condition}) guarantees that condition (a) of Proposition \ref{prop:stability} holds for all perturbed line dynamics.\\
Now, to ensure that condition (b) of Proposition \ref{prop:stability} is met for every perturbed plant $\widetilde{G}_{Lp} \in \Pi_G$, we start by noting two key points. Initially, it is important to recognize that the bound in (\ref{eq:coupling_stability_upper_bound}) is exclusively dependent on the parameter $\lambda$. Furthermore, for any given $\lambda\in \Pi_\lambda$ we have
{\small\begin{align}
    \frac{\left\|\left(\omega + j\lambda_{min}\right)^2 - \omega_0^2\right\|_2}
    {\omega_0
    \sqrt{\left(\omega + \omega_0\right)^2 + \lambda_{min}^2}}
    <
    \frac{\left\|\left(\omega + j\lambda\right)^2 - \omega_0^2\right\|_2}
    {\omega_0
    \sqrt{\left(\omega + \omega_0\right)^2 + \lambda^2}},\; 
    \forall \omega.
\end{align}}%
Therefore, the most restrictive or the smallest lower bound is associated with the case of $\lambda_{min}$ as shown in Fig. \ref{fig:lambda_stability}. Consequently, if the bound specified in (\ref{eq:coupling_stability_upper_bound}) is satisfied for $\lambda_{min}$, it will inherently be satisfied for all other potential values of $\lambda$ within the set $\Pi_\lambda$.
\end{proof}%

\begin{proof}[Proposition 4.3]
(a) We achieve zero steady-state error iff the final value condition
$   \lim_{s \to 0} \widetilde{S}
    \vv{w}s
    = 0$ in (\ref{eq: GFL_Objectives_Laplace}) is satisfied. We can rewrite the final value condition as
{\small\begin{equation}
\begin{split}
    \lim_{s \to 0} \left(
    \widetilde{S}\Gamma
    \begin{bmatrix}
    i_0^d & i_0^q
    \end{bmatrix}^{\top}
    + 
    v_0\widetilde{S}\widetilde{G}_L 
    \begin{bmatrix}
    1 & \Delta\omega_g/s
    \end{bmatrix}^{\top} \right)= 0,
\end{split}
\label{eq:ss_zero}
\end{equation}}%
where we used the disturbance type from (\ref{eq:disturbance_type}) and assumed constant reference $\vv{i_0}$. It is fairly easy to validate that (\ref{eq:ss_zero}) holds only under the given proposition. $\blacksquare$ \\
(b) We achieve synchronization or $v_c^q=0$ iff the final value condition
$   \lim_{s \to 0} [0\; K_1^q]\widetilde{S}
    \vv{w}s
    = 0,$ in (\ref{eq: GFL_Objectives_Laplace}) is satisfied.
We can rewrite the final value condition as
{\small\begin{equation}
\begin{split}
    \lim_{s \to 0}
    \begin{bmatrix}
    0 & K_1^q \widetilde{S}^q
    \end{bmatrix}
    \left(\Gamma 
    [
    i_0^d \; \; i_0^q
    ]^{\top}
    + v_0\widetilde{G}_L
    \begin{bmatrix}
    1 & \Delta\omega_g/s
    \end{bmatrix}^{\top}\right)
    &= 0.
\end{split}
\label{eq:final_value_synchronization}
\end{equation}}%
Summoning the definition of $\widetilde{S}^q$, the (\ref{eq:final_value_synchronization}) is satisfied if and only if $K_1^q/K_2^q$ has at least two zeros at the origin. $\blacksquare$ \\
(c) Based on (\ref{eq: SISO_Closed_Loop}) we have
{\small\begin{align}
    \|\vv{\hat{e}}(j\omega)\|_2
    \leq
    \|\widetilde{S}\Gamma\|_2
    \|\vv{\hat{i}_0}(j\omega)\|_2
    +
    \|\widetilde{S}\widetilde{G}_L\|_2 
    \|\vv{\hat{d}}(j\omega)\|_2,
    \;\; \forall \omega.
\end{align}}%
Therefore, the attenuation effect of the proposed closed-loop on the input set-point and grid disturbance is
{\small\begin{align}
    \frac{\|\vv{\hat{e}}(j\omega)\|_2}
    {\|\vv{\hat{i}_0}(j\omega)\|_2} &\leq
    \|\widetilde{S}(j\omega)\Gamma(j\omega)\|_2,
    &
    \frac{\|\vv{\hat{e}}(j\omega)\|_2}
    {\|\vv{\hat{d}}(j\omega)\|_2} &\leq
    \|\widetilde{S}(j\omega)\widetilde{G}_L(j\omega)\|_2.
\end{align}}%
Within the control bandwidth we have $\widetilde{S}\Gamma \approx 1/KG_L$ and $\widetilde{S}\widetilde{G}_L \approx 1/K$ therefore
{\small\begin{align}
    \frac{\|\vv{\hat{e}}(j\omega)\|_2}
    {\|\vv{\hat{i}_0}(j\omega)\|_2} &\leq
    \frac{1}{\|K(j\omega) G_L(j\omega)\|_2},
    &
    \frac{\|\vv{\hat{e}}(j\omega)\|_2}
    {\|\vv{\hat{d}}(j\omega)\|_2} &\leq
    \frac{1}{\|K(j\omega)\|_2}.
\end{align}}%
The above equation completes the proof. $\blacksquare$
\end{proof}%

\begin{proof}[Proposition 4.4]
$\hat{u}_\theta= [0\;K_2^q]\vv{e}$ and therefore using (\ref{eq: SISO_Closed_Loop})
{\small\begin{equation}
\begin{split}
    \hat{u}_\theta &=
    \begin{bmatrix}
    0 & K_2^q
    \end{bmatrix}
    \widetilde{S}
    \left(\Gamma\vv{\hat{i}_0} + \widetilde{G}_L\vv{\hat{d}}\right),
    \;\; \text{and}\;\;
    \widetilde{T}_\theta^q=K_2^q\widetilde{G}_L\widetilde{S}^q.    
\end{split}
\label{eq:inertia_control_effort}
\end{equation}}%
Finally, based on Proposition (\ref{prop:synchronization}) $K_2^q$ most include at least two integrators while $K_1^q/K_2^q$ has at least two zeros. Therefore, $\lim_{\omega \to 0} \widetilde{T}_\theta^q(j\omega) = 1$
. Moreover, $K_2^q$ is proper and $\widetilde{G}_L$ is strictly proper, hence $\widetilde{T}_\theta^q$ rolls off at high frequency, making it a low-pass filter with unity dc gain. $\blacksquare$
\end{proof}%

\begin{proof}[Proposition 5.1]
The nominal sensitivity $\widetilde{S}$ is stable with no RHP zeros; moreover, $\widetilde{G}_L K$ is of the relative degree of at least two. Hence, as a direct application of Bode's sensitivity integral, we have
{\small\begin{equation}
    \begin{split}
        \int_{0}^{\infty}\ln|\widetilde{S}
        \left(j\omega\right)|d\omega=0.
        \label{eq:Bode_Sens}
    \end{split}
\end{equation}}%
We split (\ref{eq:Bode_Sens}) into three distinct integrals over the frequency range as shown in Fig. \ref{fig:Water_Bed} and rewrite it as 
{\small\begin{equation}
    \begin{split}
        \left| \int_{0}^{\omega_{B}}\ln{|\widetilde{S}|}d\omega \right| = 
        \int_{\omega_{B}}^{\omega_{T}}\ln{|\widetilde{S}|}d\omega+
        \int_{\omega_{T}}^{\infty}\ln{|\widetilde{S}|}d\omega.
    \end{split}
    \label{eq: Bode_three}
\end{equation}}%
Both of the integrals on the right-hand side are positive, finite, and bounded
{\small\begin{align}
    \int_{\omega_{B}}^{\omega_{T}}\ln{|\widetilde{S}|}d\omega
    \leq& \left(\omega_T - \omega_B\right) \ln M_s,
    \,
    \int_{\omega_{T}}^{\infty}
    \ln{|\widetilde{S}|} d\omega \leq
    \frac{3}{4}\omega_{T},
    \label{eq:Upper_Bound_Second_Integral} 
\end{align}}%
where the first bound follows from the definition of $M_s$ and the second is based on the condition in (\ref{eq:loop_upper_bound}). A detailed proof of the second inequality is given in \cite{Seron1997}. We derive (\ref{eq:peak_sensitivity_lower_bound}) by replacing (\ref{eq:Upper_Bound_Second_Integral}) with (\ref{eq: Bode_three}). $\blacksquare$
\end{proof}%

%

\bibliographystyle{Bibliography/IEEEtranTIE}
\bibliography{Bibliography/BIB_xx-TIE-xxxx} 

\vfill

\end{document}